\pdfoutput=1

 \documentclass[final,5p,times,twocolumn]{elsarticle}

\usepackage{booktabs} 
\usepackage{tabularx}
\usepackage{tablefootnote}
\usepackage{todonotes}
\usepackage{multirow} 
\usepackage{amsmath}
\usepackage{url}
\usepackage{comment}
\usepackage{arydshln}

\usepackage{todonotes}
 \usepackage{hyperref}

\usepackage[final]{changes}  
\definechangesauthor[name={Vahid}, color=blue]{fc}
\definechangesauthor[name={Mika}, color=brown]{mm}

\newcommand{\chapquote}[3]{\begin{quotation} \textit{#1} \end{quotation} \begin{flushright} - #2, \textit{#3}\end{flushright} }

\begin{document}
\begin{frontmatter}




\title{Citations in Software Engineering --  \\Impacts of Paper-related, Journal-related, and Author-related Factors}




\author{Mika V.~M{\"a}ntyl{\"a}}
\ead{mika.mantyla@oulu.fi}
\address{M3S, ITEE, University of Oulu, Finland}

\author{Vahid Garousi}
\ead{vahid.garousi@wur.nl}
\address{Information technology group, Wageningen University, Netherlands}




\begin{abstract}
Many factors could affect the number of citations to a paper. 
Citations have an important role in research policy and in measuring the excellence of research and researchers. 
This work is the first study in software engineering (SE) to assess multiple factors affecting the number of citations to SE papers.
We use (a) negative binomial regression and (b) quantile regression to study arithmetic mean and median expected citations of a paper. Our dataset includes all the 25,113 papers which have been published in a set of 16 main SE journals, between 1970 and 2018. Our results indicate that 
publication venue, author team's past citations, paper length, the number of references, and the recency of references are the most influential factors on the number of citations to SE papers. From our empirical findings, we present several implications and advice to researchers for getting higher citations on their papers, which are in addition to the obvious case of conducting high-quality technical research, e.g. (1) Aim for high-profile venues, (2) Build a high-quality author team with highly cited past papers, and (3) Aim for high-quality work that has comprehensive content (thus longer paper length and reference list).
\end{abstract}


\end{frontmatter}



\chapquote{``If I have seen further, it is by standing on the shoulders of giants''}{Isaac Newton}{letter to Robert Hooke, 1675}

\section{Introduction}


The inspirational quote by Newton \cite{newton1675letter} reminds us of the importance of prior studies in science. In the contemporary research literature, prior studies are acknowledged by citing them. As such, past works and citations play a key role in the evolution of knowledge \cite{hamrick2010assessing}. Citations are used to document sources of information, to acknowledge prior relevant research, and to substantiate claims. However, modern and formal use of citations in scientific literature dates back only to the nineteenth century as scholars and scientists started to give continuity to their body of ideas \cite{hamrick2010assessing}. 

Citations are often used to quantify the impact of papers, journals, individuals and universties and even countries  \cite{king2004scientific} -- a practice not without controversy. 
Citation counts fall under the subject area of bibliometrics. 
Many countries are moving towards research policies that emphasize excellence; consequently, they develop evaluation systems to identify universities, research groups, and individual researchers that can be said to be 'excellent'. Such an excellence is usually measured by citation counts \cite{danell2011can}. A recent interview study \cite{ferretti2018research} reported that many scientists feel uncomfortable on using citations and other bibliometric measures to track research excellence. Yet, few alternatives and ``inescapability of quantification'' suggest that there are no easy solutions \cite{ferretti2018research}. Thus, in lack of better proxies, we measure what we can.

The subject of assessing research excellence based on citation counts has received increasing attention in science policy over the last few decades. This has led to increasing numbers of bibliometric studies, in many areas of science, to assess the factors affecting the number of citations to papers, e.g., \cite{Onodera2015, Wang2011, Tahamtan2016}. According to the findings, some of those factors include, among others  \cite{Tahamtan2016}: (1) paper-related factors such as rigor (technical quality) of paper, novelty and interest of subject, characteristics of the titles and abstracts (e.g., their length), length of paper, and characteristics of references; (2) journal-related factors such as journal impact-factor; and (3) author-related factors: such as number of authors; authors' reputation, number of and citations to their past papers. Therefore, one can argue that citation count is not a direct measure of the quality of research in a given paper, since the number of citation count to an article is influenced by various "extrinsic" factors, not directly related to its content, e.g., as listed above, author-related factors.

While some bibliometric studies have been published in SE, e.g. \cite{mathew2018finding, garousi2016citations}, only two recent studies \cite{Molléri2018, garousi2017quantity}, to the best of our knowledge, have done initial exploration of factors affecting the number of citations in SE papers (more details in Section \ref{sec:Related_Work}). Similar to the motivations as discussed in other fields, e.g., \cite{Onodera2015, Wang2011, Tahamtan2016}, we need to better characterize and understand factors affecting the number of citations to SE papers, as paper quality alone is clearly not the only influencing factor. We think understanding the other factors can help the citation-based research evaluations make more informed decisions. For example, using our results, one could make a case that if one publishes highly-cited work in a lower impact-factor journal, then such citations should be more heavily weighted in research quality evaluations. The idea of this argument lies in our finding that papers published in high-impact factor SE journals tend to be more highly cited and thus a paper from low impact-factor journal has to fight an uphill battle from the very beginning.  


The current paper uses a data set from Scopus covering all SE journal papers published between years 1970 and 2018. We extend the work reported in \cite{garousi2017quantity} by considering several additional factors, e.g., paper-related factors such as paper age, paper length (number of pages), and title length; and author-related factors such as the number of past papers, citations to past papers, and size of author team. Furthermore, as an important novelty compared to two existing related works in SE \cite{Molléri2018, garousi2017quantity}, we use multivariate analysis methods as we develop and utilize negative binomial and quantile regression models to explore the relationship between the set of citation-predictor factors and a number of citations as the response variable.

The remainder of this paper is structured as follows. Section \ref{sec:Related_Work} provides a review of the related work. Section \ref{sec:Methodology} discusses our methodology. Results are presented in Section \ref{sec:Results}. Then we discuss the results and reflect them to prior work in Section \ref{sec:discussion} Finally, in Section \ref{sec:Conclusions}, we conclude the paper, and discuss the implications of the results and also the future work directions.

\section{Related Work}
\label{sec:Related_Work}
The SE research literature has witnessed many recent bibliometric studies,  e.g., \cite{Molléri2018, garousi2016citations}. A brief survey of the bibliometric papers in SE  can be found in \cite{garousi2017quantity}.

The study reported in \cite{Molléri2018} analyzed  the relation between citations and research quality (rigor) in a set of 718 SE papers. To quantify research quality of a given paper, authors of \cite{Molléri2018} assessed its research rigor and industrial relevance, using scoring rubrics, as defined in another study \cite{ivarsson2011RigorRelevance}. The results showed that only rigor is related to studies’ normalized citations. As expected, industrial relevance of papers was not an influential factors affecting number of citations, since industrial relevance often does not often play a role when other researchers cite a given paper.

Our past work \cite{garousi2016citations} is an example of the bibliometric studies in SE, which investigated citations, research topic trends and active countries in SE, based on a pool of 72,787 papers, published in SE journals and conferences between 1972-2014. The study found that nearly half of the SE papers, in the above large dataset, were not cited at all, as of 2016.

In other areas of science (outside of SE), a large number of studies have studied factors affecting number of citations, e.g., \cite{Onodera2015, Wang2011, Tahamtan2016}.

Onodera and Yoshikane \cite{Onodera2015} used multiple regression to examine the factors affecting citation rates of papers, in six selected subject fields: (1) condensed matter physics, (2) inorganic and nuclear chemistry, (3) electric and electronic engineering, (4) biochemistry and molecular biology, (5) physiology, and (6) gastroenterology. As explanatory regression variables, the study considered: various article-related factors such as authors' achievements (measured by past publications and citedness), and impact factors of the publishing journals. The findings showed that "Price" index \cite{price1970citation} (defined below) was the strongest predictor of citations, and number of references in a given paper was the next. The "Price" index was defined as the percentage of the references cited in a given paper, whose publication year is within five years before the publication year of the citing paper. The effects of number of authors and authors' achievement measures were rather weak. 
 
Wang et al. \cite{Wang2011} used data mining to determine typical features for highly-cited papers. Their findings showed that features such as the reputation of the authors and journals, contribute to high citations.

Tahamtan et al. \cite{Tahamtan2016} reported a systematic review of the literature on factors affecting number of citations. The review study reviewed a pool of 198 primary studies. Three general categories with twenty eight factors were identified to be related to the number of citations, as we summarize in the following: 
\begin{itemize}
\item Category 1: Paper-related factors: quality of paper; novelty and interestingness of subject; characteristics of fields and study topics; methodology; paper type (conference or journal paper); study design; characteristics of results and discussions; use of figures and appendix in papers; characteristics of the titles and abstracts (such as their length); characteristics of references; length of paper; age of paper; early citation and speed of citations to papers; accessibility and visibility of papers. 
\item Category 2: Journal-related factors: journal impact factor; language of journal; scope of journal; form of publication (online or hard-print). 
\item Category 3: Author-related factors: number of authors; author's reputation; author's academic rank; self-citations; international and national collaboration of authors; authors' country; gender, age and race of authors; author's productivity; organizational features; and funding. 
\end{itemize}

The finding of the review study showed that factors such as the quality (rigour) of the paper, journal impact factor, number of authors, visibility and international cooperation are stronger predictors for citations, than characteristics of results and discussion and so on.

In SE, only two papers have explored factors affecting number of citations \cite{Molléri2018, garousi2017quantity}. Garousi and Fernandes \cite{garousi2017quantity} analyzed four possible factors: (1) paper types (conference versus journal papers), (2) publication venues; (3) authors' country of affiliations, and (4) papers written by each of the top-10 authors in SE (in terms of number of papers). The study treated each of the above factors in isolation by treating them as mutually independent and thus did not use any form of regression analysis over a set of multiple factors. 


\textit{Relationship of our paper with above two previous works in SE:}  Based on the above discussions, the current paper is the third paper in the SE discipline, after \cite{Molléri2018, garousi2017quantity}, studying the factors affecting the number of citations. The current paper extends the work reported in \cite{garousi2017quantity} by considering several additional factors, e.g., paper-related factors such as paper age, paper length (number of pages), and title length; author-related factors such as a number of past papers, citations to past papers, and size of author team. Furthermore, as a more important novelty compared to both \cite{garousi2017quantity, Molléri2018}, we develop and use a multivariate analysis, negative binomial and quantile regression, models to explore the relationship between the set of multiple citation-predictor factors and citations as the response variable. Thus, this paper is the first paper in SE to use a rigorous statistical method (multivariate analysis) to study the set of factors affecting the number of citations to SE papers.



\section{Research methodology}
\label{sec:Methodology}
In this section, we present the goal and research questions (RQs) first, and then discuss our data collection and data analysis approaches.

\subsection{Goal and research questions}

The research approach we have used in our study is the Goal, Question, Metric (GQM) methodology \cite{caldiera1994goal}. Using the GQM's goal template [38], the goal of this study is to systematically identify the factors affecting citation rates of journals papers in SE, from the point of view of the researchers in this area. Based on the above goal and by investigating the prior work in the area, we raised the following research questions (RQs). As one can see, the goal and RQs of the study are \textit{exploratory} and \textit{descriptive} in nature \cite{easterbrook2008selecting}.

\begin{itemize}
\item RQ 1: How does age of a paper (years since publications) affect its citation count?
\item RQ 2: How does the publication venue affect citation count? 
\item RQ 3: How do the author team characteristics affect citation count? 
\item RQ 4: How do the paper characteristics affect citation count? 
\item RQ 5: Is there a difference in significant factors when predicting expected mean or median citations?
\end{itemize}

We show in Table \ref{table:list_of_factors_for_RQs} the list of the factors that we planned to analyze for each of the RQs. Note that, as we discuss in Section \ref{sec:Data_collection}, for deciding which factors to include in our analysis, two main determining factors were as follows: (1) whether data (measures) for a given factor (e.g., number of co-authors) were available via the academic database that we were going to select; and (2) whether we could automatically extract the objective values of a given factor via the academic databases. Thus, we did not include "subjective" factors such as quality of paper; novelty and interest of subject; study design; characteristics of results and discussion; for which no data were already available. 

\begin{table*}[hbt!]
\centering
\caption{Factors that we analyzed for each of the RQs}
\label{table:list_of_factors_for_RQs}
\begin{tabular}{ll}
RQ                          & Factors                                                                                                                                                                                                                                        \\
\hline
\hline
RQ1: Age                    & Age (years since publication)                                                                                                                                                                                                                  \\
\hline
RQ2: Venue                  & \begin{tabular}[c]{@{}l@{}} Individual Venues \\Impact factors of the venues  \end{tabular}                                                                                                                                                                                                                  \\
\hline
RQ3: Author-related factors & \begin{tabular}[c]{@{}l@{}}Number of co-authors\\ Past citations of the authors\\ Number of authors' past papers\\ Number of different affiliations\\ Number of different countries\end{tabular}                                                  \\
\hline
RQ4: Paper-related factors  & \begin{tabular}[c]{@{}l@{}}Title length \\ Paper type (e.g., review or regular papers)\\ Paper length (number of pages)\\ Number of references in the paper\\
Recency of references\end{tabular} 
\\
\hline
RQ5: Mean or Median citations  & All factors from RQ1-RQ4 
\\
\hline
                            &                                                                                                                                                                                                                                               
\end{tabular}
\end{table*}

The paper age is studied in RQ1. The age of a paper affects the number of citations it has received since accumulation of citations to a given paper takes time. It is thus important to consider age of a paper as a control factor in all regression models. 
Another alternative to the above choice would be to predict the citations for articles within certain age groups, an analysis which would not allow to conduct regression analysis on a large set of all SE papers. 
For example, in \cite{Onodera2015}, the authors predicted citations of articles in age groups of 6 and 11 years.  Typically, the number of citation to a paper  increases faster in the early years and then slower in later years \cite{Tahamtan2016}.  
To model the above phenomenon accurately, we needed a non-linear model of age. Therefore,  we included the age factors with two terms: (1) a logarithmic age,  and (2) linear age. We also tested terms (1) linear age and (2) quadratic age but the former combination had better model quality (measured with AIC \cite{akaike1987factor} explained in Section \ref{sec:model_quality}).

Where the article is published, i.e., the publication venue, is also likely to affect its citation count (RQ2). After all, the journal impact-factors are effectively measuring the mean number of citations of the articles, published in a particular journal, receive in the course of two years after they have been published. We planned to select the top SE journals, and then study the differences in citations to papers published in them. 

Author-team characteristics (RQ3) are also likely to affect the citation count. After all, the quality of the work that authors have done previously might be repeated in the future papers or the this could be due to 'halo-effect'  as a paper by a highly cited author has a chance to get more citations just because of the authors of the paper rather than the actual paper content, an observations which has been reported in the literature  \cite{johnson1997getting, hudson2007known}.  

RQ4 analyzes paper-related characteristics such as page counts and the number of references in a paper on the number of citations, which have been discussed as influential factors in previous studies  \cite{Tahamtan2016}. Longer papers are in general expected to have more materials (and thus more contributions) and are likely to be more cited. 
The number of references in a given paper, on the other hand, denotes how thorough the authors have been in studying prior works. The higher the number of references (to more recent studies) in a paper, the higher the chance of getting more citations \cite{Tahamtan2016}. Also, there is a form of 'reciprocal altruism' w.r.t. citing and getting cited in the research community in general, when a researcher cites the papers of other researchers, s/he could more likely get citations in the papers written by other researchers \cite{corbyn2010easy}. As a paper in Nature \cite{corbyn2010easy} mentioned: ``Scientists are subject to social forces as much as anyone in any other profession", and that, ``If you want to get more cited, the answer could be to cite more people". That study analyzed over 50,000 Science papers and suggested that: ``it could pay to include more references [in a paper]".


Finally, RQ5 runs through all other RQs and studies whether different predictors (factors in Table  \ref{table:list_of_factors_for_RQs}) are significant if we predict expected (arithmetic) mean or median citations. By an initial analysis of the dataset, we found that the citation data is skewed, since while the mean number of citations per paper in the  dataset is 22.48,  the median is only 6. Thus, it is plausible that we would find different  citation-impacting factors when analyzing based on expected median citations (the citation value separating the higher half of papers from the lower half) or expected mean (the sum of citations divided by the number of papers). 


\subsection{Data collection}
\label{sec:Data_collection}
Our goal was to gather all SE journal papers in a systematic manner. Given the large number of SE papers (in 10's of thousands \cite{garousi2017quantity}), we had to automate our data collection phase. To enable that phase, we discuss next how we selected the right publication search engine, and how we selected the SE journal venues.

\subsubsection{Selection of the publication search engine}

To gather the pool of SE journal papers, we had to select a suitable publication search engine. For systematic selection of such a search engine, by reviewing the related  studies and by following the approach used in our past works, e.g., \cite{garousi2017quantity, garousi2016citations}, we devised three important selection criteria: 

\begin{enumerate}
\item The publication search engine should provide the highest quality and reliability in terms of coverage of the SE literature, i.e., including all  SE journal papers, 
\item The publication search engine should include the citation data for papers,
\item The publication search engine should provide a efficient and usable API to retrieve SE papers, and as many citation-predictor factors as possible. 

\end{enumerate}

To find the candidate publication search engines, we reviewed a number of bibliometrics studies, in SE (e.g., \cite{ mathew2018finding, garousi2016citations}), and fields other than SE. We short-listed the candidate publication search engines as follows: DBLP (\texttt{www.dblp.org}), Scopus (\texttt{www.scopus.com}), Web of Science (\texttt{www.webofknowledge.com}) and Google Scholar (\texttt{scholar.google.com}). These search engines are among the most popular search engines that researchers regularly use in bibliometrics studies. DBLP was not further considered, since it does not include citation data. In Table \ref{table:three_candidate_search_engines}, we discuss how the remaining three candidate publication search engines rate in terms of the selection criteria discussed above.


\begin{table*}[hbt!]
\centering
\caption{Assessing the three candidate publication search engines in terms of the three selection criteria}
\label{table:three_candidate_search_engines}
\begin{tabular}{p{3.5cm}p{3.5cm}p{3.5cm}p{3.5cm}p{3.5cm}}
Criteria                                           & Scopus   & Web of Science    & Google Scholar                                                             &  \\
\hline
1-Quality and reliability in terms of coverage of the SE literature & Scopus has the feature to search by ISSN number, quality and reliability of search results in terms of complete coverage can be achieved. & WoS has the feature to search by ISSN number, quality and reliability of search results in terms of complete coverage can be achieved. & No ability to search based on ISSN number &  \\
\hline
2-Including citation data                                           & Yes                                                                                                                                                                               & Yes                                                                                                                          & Yes                                                                                                                          &  \\
\hline
3-API for searching and data extraction     & Allows API queries and saving the data of all extracted papers.                                                                                                                    & Allows API queries and saving the data of all extracted papers.                                    & No official API available.       & \\

\hline
\end{tabular}
\end{table*}

In conclusion, by summarizing the outcomes with respect to our three selection criteria (Table \ref{table:three_candidate_search_engines}), we selected the Scopus publication search engine as the publication search engine from which the SE papers would be retrieved. There was no major difference between Scopus and Web of Science (both seemed suitable for our purpose), but we had some previous experience with Scopus \cite{garousi2017quantity, garousi2016citations} and that is why we selected it. Several other recent bibliometric papers published in the Nature magazine, on studying top 100 cited papers \cite{van2014top} and hyper-productive authors \cite{ioannidis2018thousands}, have also used Scopus to get their data-sets. There have been empirical studies, e.g., \cite{archambault2009comparing, falagas2008comparison}, which have compared the performance and coverage of Web of Science versus Scopus in several fields, e.g., social sciences. Some studies, e.g., \cite{abrizah2013lis}, and have found empirically that Scopus is actually better than Web of Science in certain aspects, e.g., ``larger coverage of titles'' \cite{abrizah2013lis}.

\subsubsection{Selection of SE venues}
Our goal was to include all SE journal papers in our data-set that have been published prior to 2019. To decide on which journals to consider as SE journals, we included all journals included in either of the following two recent bibliometric studies in SE \cite{karanatsiou2019bibliometric, mathew2018finding}. However, we made two exceptions. First,  \cite{karanatsiou2019bibliometric} included Journal of Software (ISSN = 1796-217X), but we found that Scopus had stopped archiving its bibliometric information since 2014 although the journal still continues to publish papers. This could possibly hint towards declining quality of that journal but it also meant that we could not include it in our analysis as the post-2014 data are not available. Second, \cite{mathew2018finding} included ACM SIGSOFT Software Engineering Notes and marked it as a journal. However, according to ACM, it is not a peer-reviewed journal, as its web-site states "Software Engineering Notes (SEN) is edited, but not refereed." \footnote{https://www.sigsoft.org/SEN/}

Our original plan was to collect data from both SE journals and conferences. We collected all journal articles easily with ISSN numbers. ISSN numbers are resilient for example to different spellings of journal names that also existed in our data. However, conferences typically do not have ISSN number but rather they have ISBNs that change with every edition of the conference. ICSE is undoubtedly the flag-ship conference of SE and it has an ISSN number (0270-5257). However, when using that number, we noticed that the number of articles varied greatly across years. For example for 2018 query with ICSE's ISSN number returned 1,036 articles while, for 2017, the query returned only 50 articles. Closer examination revealed that, in 2017, only papers in the technical track were considered "worthy" of getting the ICSE's ISSN number. In 2018, it was given for all papers including ICSE workshop and doctoral symposium papers. Due to the  difficulties in extracting conference papers from Scopus, we decided only to include journal papers.

We collected all the data, needed for our study,  from Scopus database between 12 and 15 February 2019. According to Elsevier, Scopus "is the largest abstract and citation database of peer-reviewed literature: scientific journals, books and conference proceedings" \footnote{https://www.elsevier.com/solutions/scopus}. We utilized Elsevier's web-API and used its documentation  \footnote{https://dev.elsevier.com/}. The scripts that we developed and used to fetch and analyze the data are available as a public link\footnote{https://tinyurl.com/y5jb55em} to Dropbox folder (if the paper gets accepted we will move the scripts to Github or other more permanent location as we have done in the past). 

We are happy to share the data but we must honor Scopus's policy of data sharing that states ``Public sharing of data for purpose of reproducibility with a specific party is permissible upon written request and explicit written approval.''\footnote{https://dev.elsevier.com/policy.html} 

\begin{figure}[hbt!]
  \centering
  \includegraphics[width=90mm]{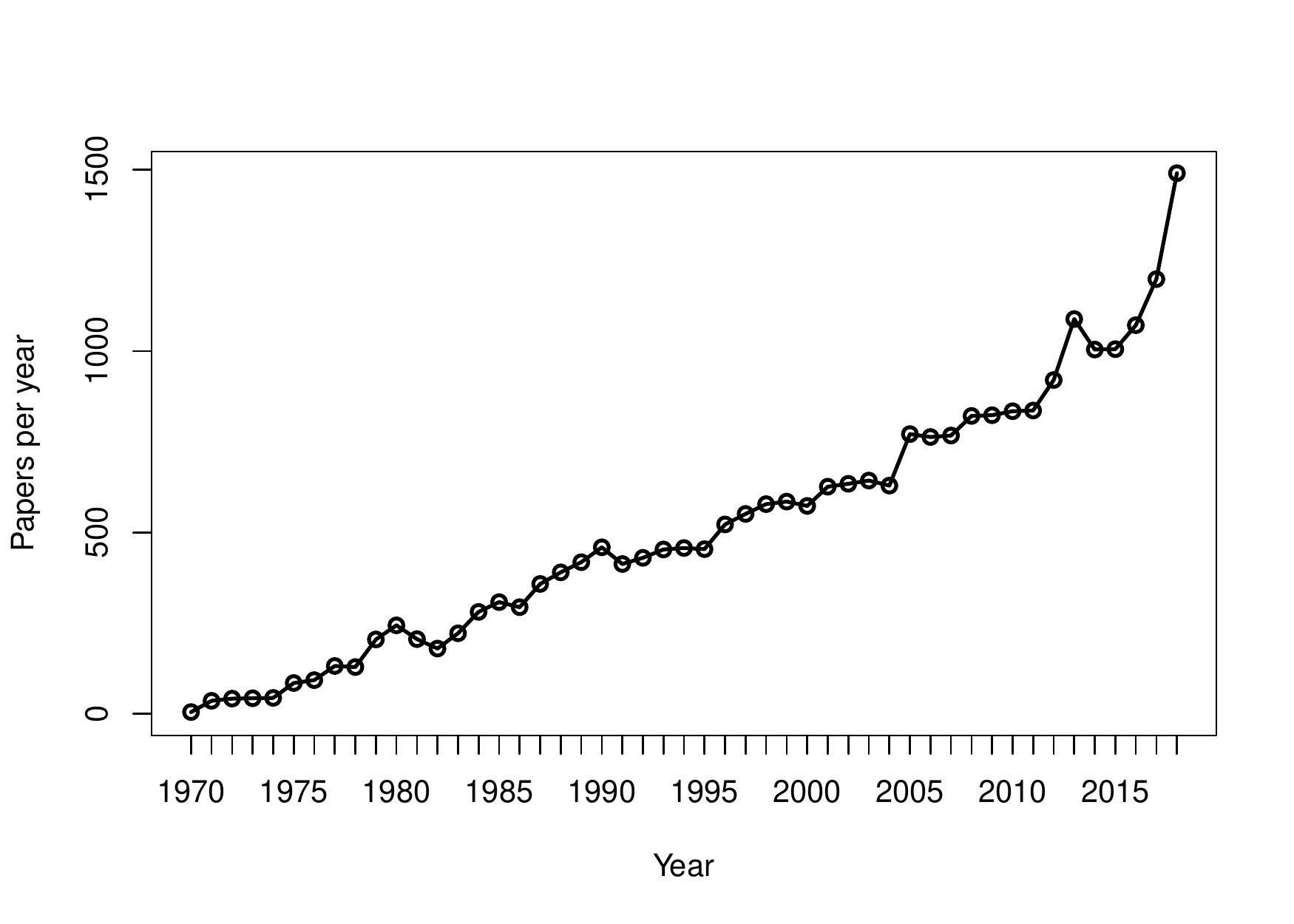}
  \caption{Number of software engineering journal papers published per year}
  \label{fig:PapersPerYear}
\end{figure}

After conducting our paper search activity, we ended up with 25,113 papers, published in 16 SE journals. Table ~\ref{tab:journals} shows all the journals as well as the number of papers published in each journal. We can see that the five journals, based on the number of papers, in this list are: JSS, TSE, SWPE, IEEESW, and IST. These five journals include 72\% (18,076 / 25,113) of the papers in the dataset. Figure ~\ref{fig:PapersPerYear} shows the number of papers per year in our data, and we can see constant growth starting from 1970 with 5 papers per year to 1,490 papers in 2018. 

\begin{table*}[hbt!]
\centering
\caption{Number of papers published in each journal, in the dataset from Scopus}
\label{tab:journals}
\begin{tabular}{rllr}
  \hline
 & Short name & Journal name & Count \\ 
  \hline
1 & ASE & Automated Software Engineering & 459 \\ 
  2 & ESE & Empirical Software Engineering  & 842 \\ 
  3 & IJSEKE & International Journal of Software Engineering and Knowledge Engineering & 1,115 \\ 
  4 & ISSE & Innovations in Systems and Software Engineering & 337 \\ 
  5 & IST & Information and Software Technology & 2,943 \\ 
  6 & JSEP & Journal of Software: Evolution and Process / Journal of Software Maintenance and Evolution & 765 \\ 

  7 & JSS & Journal of Systems and Software  & 4,483 \\ 
  8 & JSTT & International Journal on Software Tools for Technology Transfer & 728 \\ 
  9 & REJ & Requirements Engineering & 408 \\ 
  10 & SOSYM & Software and Systems Modeling & 730 \\ 
  11 & SQJ & Software Quality Journal & 701 \\ 
 12 & STVR & Software Testing Verification and Reliability & 484 \\ 
  13 & SW & IEEE Software & 3,494 \\ 
  14 & SWPE & Software - Practice and Experience & 3,472 \\ 

 15 & TOSEM & ACM Transactions on Software Engineering and Methodology & 468 \\ 

  16 & TSE & IEEE Transactions on Software Engineering & 3,684 \\ 
   \hline
\end{tabular}
\end{table*}

\subsection{Regression Models}
\label{sec:data-analysis}
Let us recall that our goal is to understand what factors affect citations to SE papers. A literature review on factors affecting citations \cite{Tahamtan2016} pointed that using univariate analysis such as correlation or t-tests provides weaker evidence than multivariate analysis as the latter can reveal also interactions between variables. From \cite{Tahamtan2016,Onodera2015}, we can find that two most prevalent methods for analyzing citations are: (1) regular linear regression with the independent variable transformed to its log or square root, and (2) negative binomial regression. 


Out of these two alternatives, negative binomial regression \cite{hilbe2011negative} is better as it does not require transforming the independent variables to meet the requirement of the normal distribution that is precursory of independent variables in linear multiple regression. Furthermore, negative binomial regression is designed for count data where there are no negative values. 

Negative binomial regression is designed to model negative binomial distribution. 
According to Hilbe \cite{hilbe2011negative}, this distribution is given by:
\begin{equation*}
 \Pr(Y = y) = \frac{\Gamma(1/\alpha+y)}{\Gamma(y+1) \, \Gamma(1/\alpha)} \left(\frac{1}{1+\alpha\mu}\right)^{1/\alpha} \left(\frac{\alpha\mu}{1+\alpha\mu}\right)^y 
\end{equation*}

Here $\mu$ is the mean of Y (Y is citations in our case) and $\alpha$ is the heterogeneity parameter. The expected value  is evaluated with generalized linear model 
\begin{equation*}
ln(\mu) = \beta_0 + X_1\beta_1 + ... + X_k\beta_k
\end{equation*}

Negative binomial regression optimizes the regression model towards minimizing the sum of squared errors using the so-called Iteratively Reweighted Least Squares (IRLS) method \cite{holland1977robust}. The value that minimizes the sum of squared errors is the mean expected citations of an article.
We use negative binomial regression as our analysis method to predict mean citations. We used the implementation in R-package MASS  \cite{ripley2013package}.

However, our citation data is skewed with mean=22.48 and median=6 and we wanted to study both measures (median and mean) in the regression model.  To predict median we use quantile regression \cite{koenker1978regression} from the R-package quantreg \cite{koenker2012quantile} to model median citations. 
In quantile regression, the goal is to find a regression line: \begin{equation*}
Q_\tau(Y) = \beta_0(\tau) + X_1\beta_1(\tau) + ... + X_k\beta_k(\tau) 
\end{equation*}
Here $\tau$ is the chosen quantile which in our case is the median $\tau$=0.5. The betas for the above equation is are searched by minimizing the following:
\begin{equation*}
\hat{\beta_{\tau}}=
{\mbox{arg min}}\sum_{i=1}^{n}(\rho_{\tau}(Y_{i}-X_{i}\beta))
\end{equation*}

In other words, predicting median ($\tau$=0.5) in quantile regression optimizes the regression model towards minimizing the sum absolute errors rather than the sum squared errors \cite{koenker1978regression}. 
For any other quantile, it minimizes to sum of weighted absolute errors with $\tau$ determining the weighting. The minimization is a linear programming problem which can be solved with simplex algorithm. Quantile regression does not assume a particular distribution, e.g. normal or negative binomial \cite{cade2003gentle}. It is also robust to outliers, but needs a large data set and is computationally intensive. 


Quantile regression has not been among the traditional methods in citation analysis but we found that the method is being used in the more recent articles in particular in those published in the last three years \cite{danell2011can,wang2018relationship, ahlgren2018exploring, sienkiewicz2016impact, stegehuis2015predicting}. 
Therefore, we use both regression analysis approaches, i.e., negative binomial regression to model the expected mean citations and quantile regression to model the median citations. 
\subsection{Model Quality}
\label{sec:model_quality}
In this paper, we build and compare several regression models, based on the above two regression analysis approaches, with a different number of predictor variables. 
Therefore, we need to compare model quality, e.g., if we add a new factor to the regression model, we need to assess whether the model get better or worse. To do this, we use the Akaike Information criterion (AIC) \cite{akaike1987factor}.

With our regression models, we have a trade-off between overfitting and underfitting the data. 
The AIC measure tries to balance this trade-off. AIC prefers a model with a high goodness fit and a low number of predictors. Smaller values of AIC are better and it is computed as follows $AIC=2k-2ln(L)$. Where k is the number of predictors and L is the maximum likelihood measuring the goodness of fit of the model. 

For developing regression models to answer RQ2-RQ5, our approach was to start with a model including only the age of the article (studied in RQ1). We then planned to add the other factors one-by-one starting from the one we considered to be the most likely to improve the model quality (as measured by the AIC metric). 
Our principle was to keep the newly-added factor (variable) in the model if the AIC improved and  to exclude it, if the  AIC did not improve. 



\section{Results}
\label{sec:Results}
\subsection{RQ1: Impact of paper age on citations}

\label{sec:ResRq1}

\begin{figure}
  \centering
  \includegraphics[width=\columnwidth]{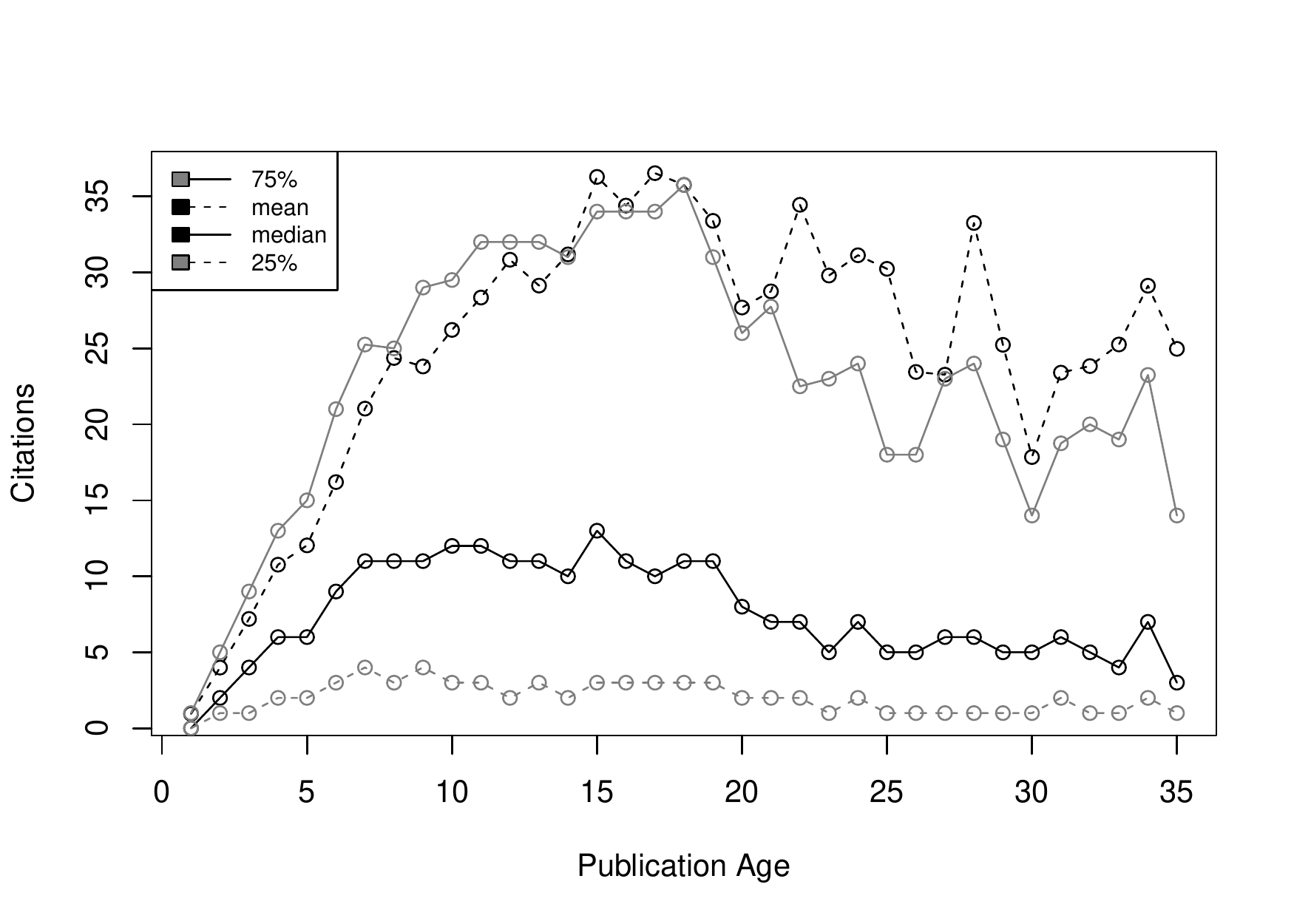}
  \caption{Citations per age (Age 1 means publication year 2018)}
  \label{fig:CitesPerAge}
\end{figure}

We first investigated the number of citations for a given age of papers (based on their publication years). Figure \ref{fig:CitesPerAge} shows mean, median, 75 percentile and 25 percentile of citations for publications of a particular age (Age 1 means the publication year 2018). The figure shows that, for the case of the first seven years in the chart (corresponding to publication years 2012-2018), citation counts increase in all measures. After 2012, the papers at and below the 25 percentile receive no more citations, in general. 
Furthermore, we can see that the increase in citation median also plateaus. We can see a "shoulder" in the Figure \ref{fig:CitesPerAge} at age 7 (2012), and median receives a higher number of citations only at age  15 (2004), 11 (2008), and 10 (2009). The 75 percentile curve plateaus after 11 years (2008-2018), although, there is a slight increase in citations until age 17 (2001). The mean citations per paper increase until age 14 (2004). 

For all variables, we can see first stark increase, which then plateaus and then turns to slight decay. This all seems intuitive excluding the finding that a number of citations decay with age. After all, once a citation has been made, it should never go down. This phenomenon is explained by the fact that the number of references used in a single paper goes up over time. In 2000, the average number of references used in a paper in our data was 10.5, while in 2018 it was 29.3. As most citations are made in few years after a paper is published, this explains why old papers have fewer citations. 

\begin{figure}
  \centering
  \includegraphics[width=\columnwidth]{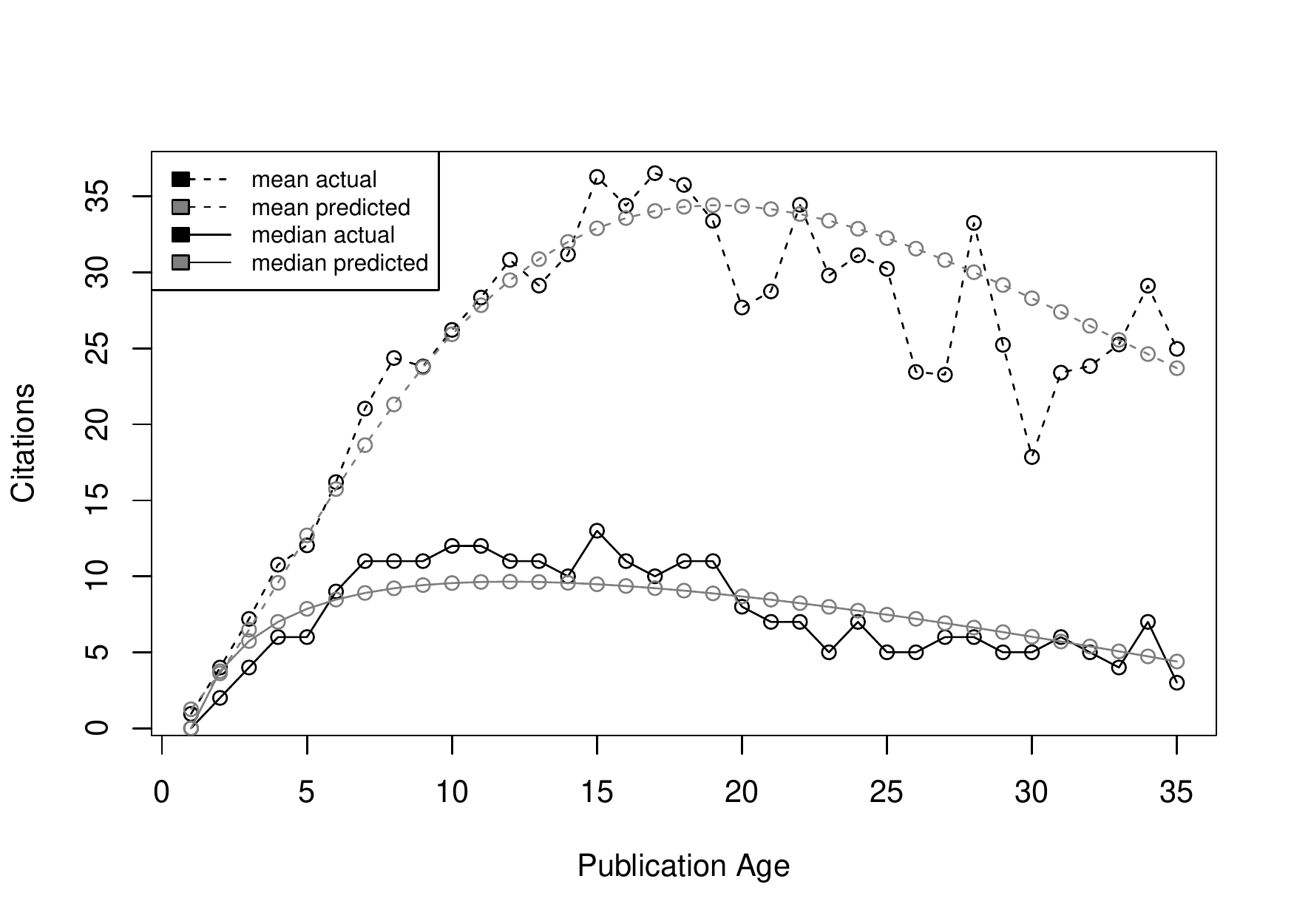}
  \caption{Mean and median citations and fitted prediction lines from NB and quantile regression models}
  \label{fig:CitesPerAgePred}
\end{figure}

To model this behavior, we use negative binomial (NB) regression for the mean and quantile regression for the median number of citations.  Figure \ref{fig:CitesPerAgePred} shows our predictions and the regression coefficients are in Tables \ref{tab:nbr_coef_age} and \ref{tab:quant_coef_age}. 


Both models and coefficients are highly significant ($p<0.0001$).  Both models have two components, one responsible for start increase at the beginning ($log(Age)$ with positive coefficient) and then a component that causes the plateau and decay (Age with negative coefficient). For both models, using use $log(age)$ and age rather than $age$ and $age^2$ resulted in better AIC scores. For binomial negative regression, AIC improved from 188,170 to 186,298 and quantile regression AIC improved from 236,455 to 235,909. 


To summarize, for low cited papers at 25 percentile, most citations occur during their first 7 years. For papers that receive more citations, the accumulation of citation occurs roughly in the first 15 years. This has quite an important implication, as it means that the impact-factor measure that usually considers only the first two years after publication is not well suited for SE articles.  


\begin{table}
\caption{Negative binomial regression model coefficients}
\label{tab:nbr_coef_age}
\centering
\begin{tabular}{lrrrr}
  \hline
Coefficients:  & Estimate & Std. Error & z value & Pr($>$$|$z$|$) \\ 
  \hline
(Intercept) & 0.3114 & 0.0330 & 9.43 & 0.0000 \\ 
  Age & -0.0865 & 0.0020 & -43.50 & 0.0000 \\ 
  Age\_log & 1.6542 & 0.0235 & 70.43 & 0.0000 \\ 
   \hline
\end{tabular}
\end{table}


\begin{table}
\caption{Quantile regression model coefficients}
\centering
\label{tab:quant_coef_age}
\begin{tabular}{lrrrr}
  \hline
Coefficients: & Value    & Std. Error & t value   & Pr($>$$|$t$|$) \\
  \hline
(Intercept)   & 0.5154  & 0.5004    & 1.0300   & 0.30302               \\
Age           & -0.5154 & 0.0221    & -23.2836 & 0.00000               \\
Age\_log      & 6.1648  & 0.3432    & 17.9644  & 0.00000              \\
  \hline
\end{tabular}
\end{table}



\subsection{RQ2: Impact of publication venues on citations}
\label{sec:ResRq2}

\begin{figure*}
  \centering
  \includegraphics[width=\textwidth]{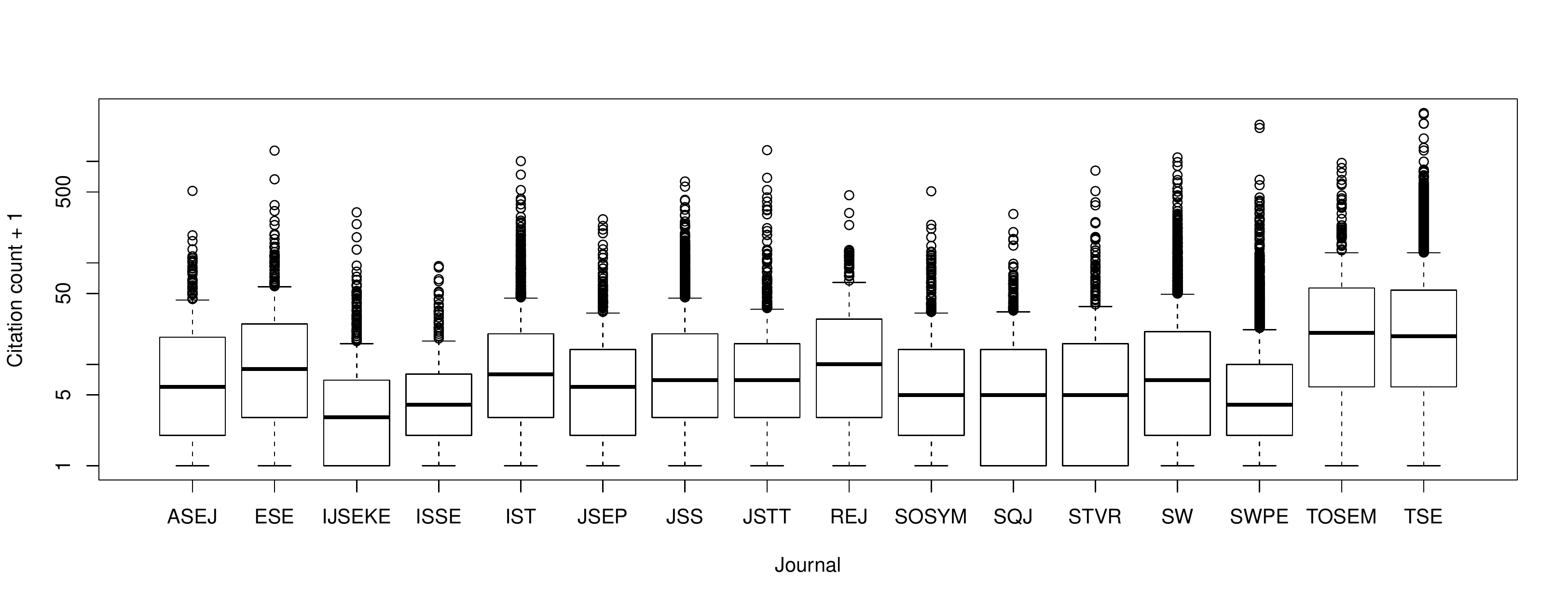}
  \caption{Citations per Journal (Note: Y-axis shows citations count + 1 as log of zero is undefined)} 
  \label{fig:CitesPerJournal}
\end{figure*}

\textbf{Individual venues.} 
Figure \ref{fig:CitesPerJournal} shows the box-plots of citations received by articles in all SE journals. The figure shows that TSE and TOSEM are the top while IJSEKE is at the bottom of the list, based on the citation count per article. However, let us note that this figure does not consider the joint effect of the age of the article and its venue. We investigate that next. 

Table \ref{tab:c_bnr_age_veneu} and Table \ref{tab:c_quan_age_veneu} show coefficients for NB regression and quantile regression when considering both age and venue. Note that when modeling a dichotomous variable in a regression model, such as a journal where a paper was published, this needs to be done with dummy variables \cite{hardy1993regression}. These dummy variables are numerical representations of a dichotomous fact. For example, in our data, a paper published in JSS has a dummy variable J:JSS set as 1 and all other dummies J:ESE, J:IJSEKE, etc, are zero. Likewise for other journals. However, this approach requires that one dummy variable is set as the default case. ASE journal is set as the default case to which all other venues are compared against and does not appear in the table. This means that for ASE papers all dummy variables are zero.   

\begin{figure}
  \centering
  \includegraphics[width=\columnwidth]{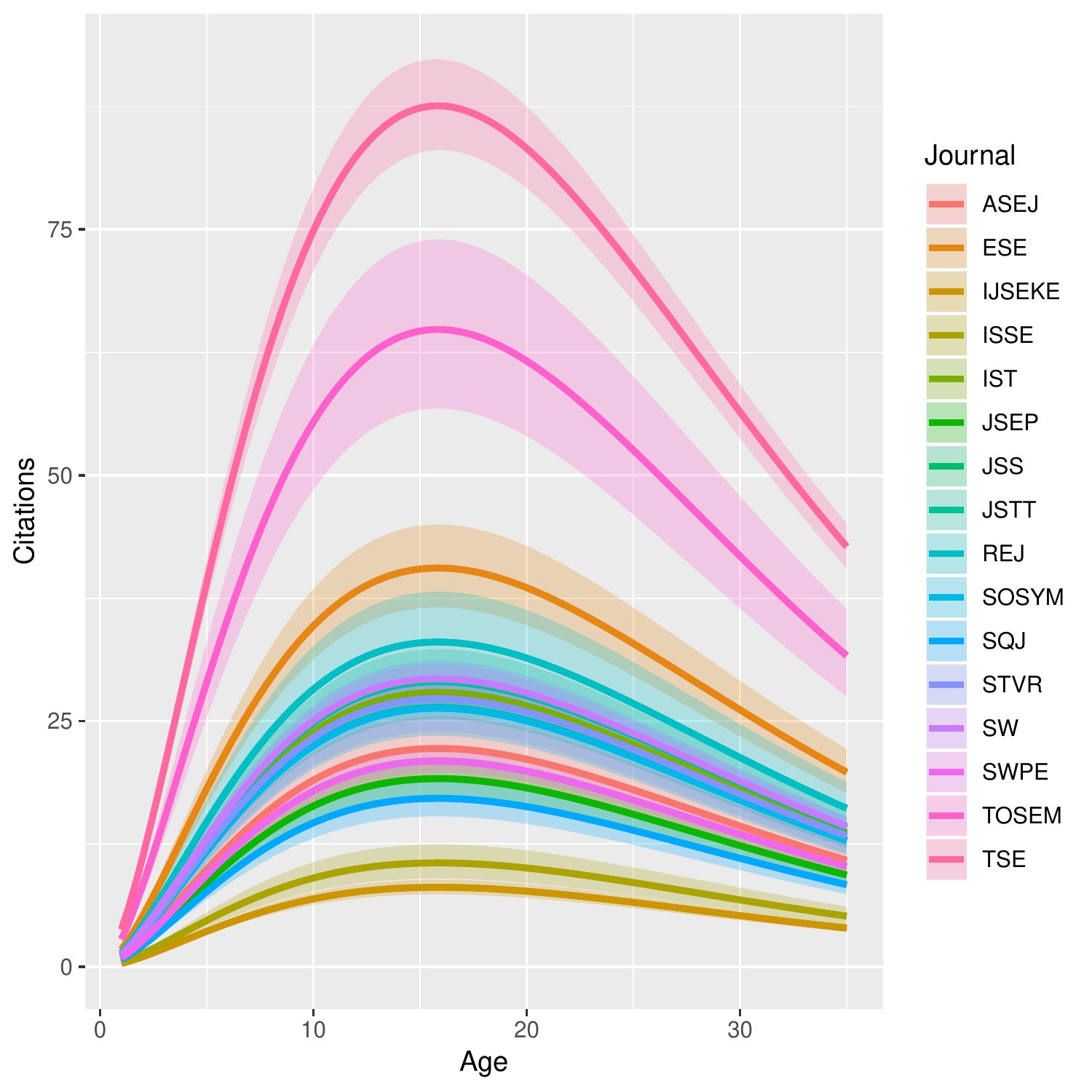}
  \caption{Citations and 95\% confidence interval by age with negative binomial regression}
 \label{fig:CitesPerJournalPerYearNB}
\end{figure}

\begin{figure}
  \centering
  \includegraphics[width=\columnwidth]{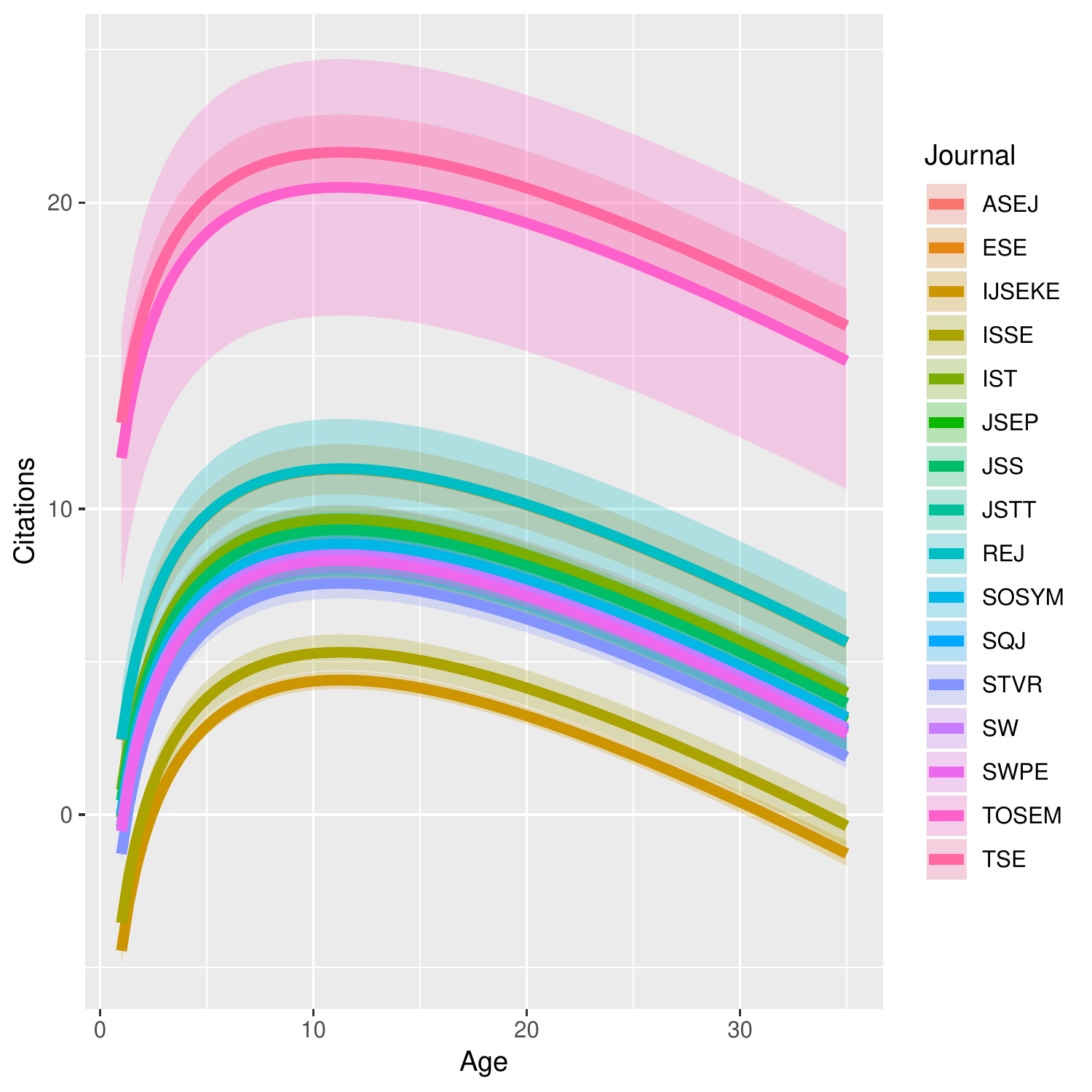}
  \caption{Citations and 95\% confidence interval by age with quantile regression}
 \label{fig:CitesPerJournalPerYearQuan}
\end{figure}

Figures \ref{fig:CitesPerJournalPerYearNB} and \ref{fig:CitesPerJournalPerYearQuan} show NB regression, for mean citations, and quantile regression, for median citations, over time for different  journals. The figures depict the line of predicted citations and 95\% confidence intervals.  From both figures, we see that TSE is associated with the highest citations in mean and median when age is part of the regression model. TSE is in both models followed by TOSEM while the third place goes to ESE and REJ, in each of the two figures. IJSEKE journal is associated with the lowest citation count in both regression models.  

\begin{table}
  \caption{Coefficient of binomial negative regression for age and venue}
  \label{tab:c_bnr_age_veneu}
\centering
\begin{tabular}{lrrrr}
  \hline
 & Estimate & Std. Error & z value & Pr($>$$|$z$|$) \\ 
  \hline
(Intercept) & 0.0720 & 0.0760 & 0.95 & 0.3433 \\ 
  Age & -0.1086 & 0.0020 & -53.02 & 0.0000 \\ 
  log(Age) & 1.7191 & 0.0231 & 74.36 & 0.0000 \\ 
  J:ESE & 0.6018 & 0.0863 & 6.97 & 0.0000 \\ 
  J:IJSEKE & -1.0092 & 0.0833 & -12.12 & 0.0000 \\ 
  J:ISSE & -0.7419 & 0.1080 & -6.87 & 0.0000 \\ 
  J:IST & 0.2282 & 0.0744 & 3.07 & 0.0021 \\ 
  J:JSEP & -0.1488 & 0.0881 & -1.69 & 0.0911 \\ 
  J:JSS & 0.1713 & 0.0726 & 2.36 & 0.0183 \\ 
  J:JSTT & 0.2667 & 0.0882 & 3.02 & 0.0025 \\ 
  J:REJ & 0.3966 & 0.1006 & 3.94 & 0.0001 \\ 
  J:SOSYM & 0.1689 & 0.0892 & 1.89 & 0.0583 \\ 
  J:SQJ & -0.2595 & 0.0896 & -2.90 & 0.0038 \\ 
  J:STVR & 0.2053 & 0.0963 & 2.13 & 0.0330 \\ 
  J:SW & 0.2766 & 0.0736 & 3.76 & 0.0002 \\ 
  J:SWPE & -0.0597 & 0.0747 & -0.80 & 0.4240 \\ 
  J:TOSEM & 1.0706 & 0.0964 & 11.10 & 0.0000 \\ 
  J:TSE & 1.3713 & 0.0740 & 18.53 & 0.0000 \\ 
   \hline
\end{tabular}
\end{table}

\begin{table}
  \caption{Coefficient of quantile regression for age and venue}
  \label{tab:c_quan_age_veneu}
\begin{tabular}{lrrrr}
 \hline
         & Value    & Std.Erro & t-value   & Pr($>$$|$t$|$) \\
         \hline
(Intercept)          & 0.51849  & 0.67082  & 0.77291   & 0.43958               \\
Age                  & -0.51849 & 0.01445  & -35.87568 & 0.00000               \\
log(Age)             & 5.85078  & 0.14416  & 40.58425  & 0.00000               \\
J:ESE    & 2.44454  & 0.76792  & 3.18335   & 0.00146               \\
J:IJSEKE & -4.44315 & 0.66910  & -6.64049  & 0.00000               \\
J:ISSE   & -3.53697 & 0.71777  & -4.92769  & 0.00000               \\
J:IST    & 0.81626  & 0.67966  & 1.20098   & 0.22977               \\
J:JSEP   & -0.09670 & 0.82822  & -0.11675  & 0.90706               \\
J:JSS    & 0.46303  & 0.68569  & 0.67528   & 0.49951               \\
J:JSTT   & 0.00000  & 0.83916  & 0.00000   & 1.00000               \\
J:REJ    & 2.46303  & 1.05175  & 2.34184   & 0.01920               \\
J:SOSYM  & 0.00000  & 0.82290  & 0.00000   & 1.00000               \\
J:SQJ    & -0.44315 & 0.73989  & -0.59893  & 0.54922               \\
J:STVR   & -1.28076 & 0.69027  & -1.85545  & 0.06354               \\
J:SW     & -0.34253 & 0.71050  & -0.48210  & 0.62974               \\
J:SWPE   & -0.53697 & 0.67478  & -0.79578  & 0.42617               \\
J:TOSEM  & 11.65747 & 2.23423  & 5.21767   & 0.00000               \\
J:TSE    & 12.80316 & 0.90048  & 14.21808  & 0.00000              \\
 \hline
\end{tabular}
\end{table}

\textbf{Impact Factors of Venues.} Scopus offers three different impact-factor metrics for journals, namely: SJR (SCImago Journal Rank), SNIP (Source Normalized Impact per Paper) and CiteScore. SJR is designed for citation networks and it has been inspired by Google's page rank algorithm \cite{gonzalez2010new}. SNIP tries to normalize for field-specific differences between journals and it is further described in \cite{moed2010measuring}. CiteScore is Scopus's version of the traditional impact-factor measure and it considers the number of citations made in the past three years, rather than two years. 

As the metrics were not available for all years, we first computed the mean scores of all metrics and assigned them to each journal for all years as in Table \ref{tab:if}. For example, all TSE papers received the same CiteScore. We then tested both regression models (2) with all set of impact factors as predictors (7 of them), which result in  14 models in total. As the data revealed, shown in Table \ref{tab:c_nb_age_if}, the best model in terms of AIC for NB regression is the one with CiteScore and SNIP. 
For quantile regression, the best model is with all three impact factors, see Table \ref{tab:c_quan_age_if}. 


\begin{table}[hbt]
  \caption{Average impact factor measures from Scopus}
  \label{tab:if}
\centering
\begin{tabular}{rlrrr}
  \hline
 & PubName & SNIP\_avg & SJR\_avg & CiteScore\_avg \\ 
  \hline
1 & ASE & 1.60 & 0.57 & 2.29 \\ 
  2 & ESE & 1.91 & 0.69 & 3.48 \\ 
  3 & IJSEKE & 0.66 & 0.26 & 0.62 \\ 
  4 & ISSE & 1.04 & 0.29 & 0.89 \\ 
  5 & IST & 1.72 & 0.58 & 3.39 \\ 
  6 & JSEP & 1.18 & 0.32 & 1.23 \\ 
  7 & JSS & 1.54 & 0.53 & 2.82 \\ 
  8 & JSTT & 1.69 & 0.64 & 1.65 \\ 
  9 & REJ & 2.25 & 0.63 & 2.69 \\ 
  10 & SOSYM & 2.74 & 0.89 & 1.97 \\ 
  11 & SQJ & 1.10 & 0.34 & 1.53 \\ 
  12 & STVR & 1.75 & 0.63 & 2.06 \\ 
  13 & SW & 2.19 & 0.66 & 1.78 \\ 
  14 & SWPE & 1.34 & 0.48 & 1.84 \\ 
  15 & TOSEM & 3.85 & 1.30 & 3.44 \\ 
  16 & TSE & 4.12 & 1.39 & 5.14 \\ 
   \hline
\end{tabular}
\end{table}

\begin{table}[hbt]
  \caption{Coefficient of negative binomial regression for age and impact factor}
  \label{tab:c_nb_age_if}
\centering
\begin{tabular}{lrrrr}
  \hline
 & Estimate & Std. Error & z value & Pr($>$$|$z$|$) \\ 
  \hline
(Intercept) & -0.7921 & 0.0377 & -21.00 & 0.0000 \\ 
  Age & -0.1083 & 0.0019 & -56.32 & 0.0000 \\ 
  log(Age) & 1.7191 & 0.0226 & 76.18 & 0.0000 \\ 
  CiteScore\_avg & 0.1679 & 0.0128 & 13.11 & 0.0000 \\ 
  SNIP\_avg & 0.3440 & 0.0160 & 21.44 & 0.0000 \\ 
   \hline
\end{tabular}
\end{table}

\begin{table}[hbt]
  \caption{Coefficient of quantile regression for age and impact factor}
  \label{tab:c_quan_age_if}
\begin{tabular}{lrrrr}
\hline
            & Value   & Std. Error & t value  & Pr($>$$|$t$|$)  \\
            \hline
(Intercept)  & -7.4480 & 0.3258     & -22.8590 & 0.0000                \\
Age          & -0.5204 & 0.0148     & -35.2492 & 0.0000                \\
log(Age)     & 6.1755  & 0.1514     & 40.7826  & 0.0000                \\
CiteScore\_avg & 1.3500  & 0.1299     & 10.3955  & 0.0000                \\
SNIP\_avg      & 1.2378  & 0.6397     & 1.9351   & 0.0530                \\
SJR\_avg       & 4.3535  & 2.2360     & 1.9470   & 0.0515  \\
\hline
\end{tabular}
\end{table}

The model with the impact-factor metrics received weaker AIC than the model with journal names as dummy variables. 
The AIC value in negative binomial regression when using journal names as dummy variables was 183,093 (details in Table \ref{tab:c_bnr_age_veneu}). When we used all three impact-factor measures, we get a lower AIC  value (183,281), and details can be found in Table \ref{tab:c_nb_age_if} . 
AIC in quantile regression when using journal names as dummy variables was 234,598 and when we use all three impact-factor measures we get a lower AIC  (234,724). Of course, the weakness of journal names as dummy variables is that it does not scale when one has multiple venues as each venue is modelled by a new factor. 
We could not combine the journal names as dummy variables model with three averaged impact-factor models as it resulted in singularities as all journal papers had the same respective average score.  

Yearly impact-factor data  was available in Scopus for limited years only, SNIP and SJR (years 1996-2017) and CiteScore (years 2011-2017). We studied the effect of using yearly impact factors for years 2011-2017, for which a subset of 7,038 papers is available. The result with this data set was that both NB and quantile regression models, the average CiteScore was the best individual predictor resulting in the best AIC.  Thus, including yearly impact factors did not improve the result in this 7-year time frame. We were surprised by this result and also investigated years 1996-2017 with respect to yearly and average SNIP and SJR. For negative binomial regression, average SJR produced the best model. However, in quantile regression, yearly SJR produced the best model. 

Given that the yearly impact factors do not improve AIC, and require excluding a very large number of papers due to missing values, we do not use them in the complete model. 

Coefficients of NB regression for all impact-factor scores is shown in Table \ref{tab:c_nb_age_if_all}. We can see that half of the impact-factor coefficients are positive while the other half are negative. It is hard to reason about this as understanding the differences between the impact-factor scores is not evident. 

Overall, it appears that the simpler metric CiteScore is better predictor the complex metrics SJR and SNIP in the model although the differences are not big. 
Furthermore, using the same average impact factor for all papers of each journal is favoured in 3 models while using yearly impact factors is favoured in one model only.

\begin{table}
  \caption{Coefficient of negative binomial regression for age and all impact factor scores}
  \label{tab:c_nb_age_if_all}
\centering
\begin{tabular}{lrrrr}
  \hline
 & Estimate & Std. Error & z value & Pr($>$$|$z$|$) \\ 
  \hline
(Intercept) & -0.2032 & 0.1030 & -1.97 & 0.0485 \\ 
  Age & 0.0712 & 0.0411 & 1.73 & 0.0829 \\ 
  log(Age) & 0.8139 & 0.1955 & 4.16 & 0.0000 \\ 
  CiteScore\_avg & 0.6252 & 0.0547 & 11.42 & 0.0000 \\ 
  SNIP\_avg & 0.2989 & 0.1251 & 2.39 & 0.0168 \\ 
  SJR\_avg & -1.1809 & 0.3820 & -3.09 & 0.0020 \\ 
  CiteScore\_yrl & -0.1614 & 0.0669 & -2.41 & 0.0159 \\ 
  SNIP\_yrl & -0.1681 & 0.0761 & -2.21 & 0.0273 \\ 
  SJR\_yrl & 0.5818 & 0.1660 & 3.50 & 0.0005 \\ 
   \hline
\end{tabular}
\end{table}

\subsection{RQ3: Impact of author-related factors on citations}
\label{sec:rq3}

It seems plausible that author team characteristics would be associated with an article's citation count. To get each author team's past output, we measured the number of past papers and number of papers citing those publications.  We only considered the number of papers of the author team and the citing papers. We counted the past papers of all authors, e.g. if authors A and B have both published 40 papers but they have jointly worked in 20 papers, then the author team A and B have joint output of 60 papers (20 independent for A and 20 independent for B plus the 20 papers they have jointly produced). 

In our dataset, we had 1,538 (of all 25,113, about 6\%) articles where the author team had not previously published any papers. Likewise, we had 2,695 articles where the author team had zero past citations. 

On the high end, we found an author team, authors of \cite{li2012improving}, who have published 2,729 articles. Such a high number looked extra-ordinary to us. We found in the past literature that Scopus had difficulty in disambiguation of some authors with common names from each other which could, in some cases, result in extremely high paper counts \cite{ioannidis2018thousands}. The past work also mentioned that Scopus had improved author classification in 2013 and we also noticed this in our manual checks. 

We checked authors' teams having a high number of previous papers and found many names originating from China. The author teams that had over 800 past papers or where past paper count per author was over 200 were manually checked and accounts IDs with multiple authors were identified.  
We identified 26 author IDs that contained contributions from multiple authors with same name. According to Scopus data, these 26 authors had jointly produced a massive thirty thousand papers in the entire Scopus data. However, when we removed all of their articles from our dataset, which is only a subset of Scopus data, this resulted in the removal of only 91 papers.


After this cleanup, the most published author team at the time of the publication had 1,273 past papers and it was a paper with 24 authors and from 22 affiliations  \cite{fernandez2017naming}. Additionally, at this point, we removed 159 papers due to having no author information. 

We built the author-team regression model while controlling for age (RQ1)
However, we excluded the venue (RQ2) as we wanted to compare the author-related factors model against the venue model. Let us remind that the final model in RQ5 combines all the factors. 

\textbf{Past Citations} of the authors are expected to influence citations counts by acting as a proxy of author quality and by famous author enjoying the benefits of halo effect, as it has been reported in the literature already \cite{johnson1997getting, hudson2007known}. However, we also suspect that the impact of past citations has diminishing returns, e.g., it has more effect to move from zero citations to one hundred citations than it is to move from 1,000 to 1,100 citations. 

Therefore, we log-transformed the past citations.  Note, we also tested the non-transformed citations and it did indeed performed worse than the log-transformed one measured with AIC score. 
We found that past citations are a highly significant positive predictor (p$<$0.0001) in both NB and quantile regression models when paper age is included in the model. Compared to the previous model including paper age only, in Section \ref{sec:ResRq1}, the regression model quality improves in NB (AIC from 184,908 to 183,570) and Quantile (AIC from 233,813 to 233,390).

\textbf{Number of Past Papers} of the author team can also influence citations even after controlling for past citations. This might be because the high number of past citations might have be gained by one or only a few articles only. 
For the reason of diminishing returns again, as outlined above, we log-transformed this variable as well. We found that, when controlling for past citations and papers' age, the number of past articles by a team of authors is significant in the NB model (p=0.0008) but, surprisingly, has a negative effect.  This means that an author team with a high number of past citations but with a low number of past papers provides maximal citation count. With quantile regression, the effect is positive but not significant (p=0.27630). Model quality improvement in the NB regression model is small (AIC from 183,570 to 183,562), while quantile regression model quality slightly decreases (AIC from 233,390 to 233,392). At this point, the past papers seems to have negative or no statistically significant effect. 

\textbf{Size of Author team} can also have a positive effect on citations via increased quality and exposure of the published work. 
Again, we tested both normal and log-transformed author count and found that the log-transformed variable produced a better model (smaller AIC). 
When we add author count to the model, the prediction quality slightly improves for NB (AIC changing from 183,570 to 183,399) and also for the case of Quantile regression (AIC from 232,278 to 232,062). Furthermore, the author-count factor is highly significant (p$<$0.0001) in both NB and quantile regression models. Surprisingly, the number of past articles also becomes more significant to (p$<$0.0001) from (p=0.0008) in NB regression. In quantile regression, the same happens as the number of past articles becomes highly significant (p$<$0.0001) and also changes its sign and becomes a negative predictor of citations. 

\textbf{Number of different affiliations} can improve citations beyond author count due to larger 'network effect' of author team working in multiple affiliations, an observation which has already been reported in the literature \cite{journal.pone.0033339, journal.pone.0099502}. For this analysis, we had to remove 975 papers from the dataset that had no affiliation information. 
Log transformation of affilation count did not improve AIC.  
We found that the affiliation count was significant in NB (p$<$0.0001) and in the quantile (p=0.029) regression models. The model quality improved for NB (AIC 178,320 to 178,306) and quantile (from  225,111 to 225,103). Note that, due to removing 975 articles, the AIC score is not comparable with the previous step (size of author team). 

\textbf{Number of different countries} that authors represent (in terms of their work affiliations) can also increase the citation counts due to network effects \cite{journal.pone.0033339, journal.pone.0099502}. We found 1,282 articles with no country information which we removed. Our analysis showed that the number of countries is a significant and positive predictor in NB (p=0.0041) but negative and not significant in Quantile regression model (p=0.146). Model quality improves for NB from (from 176,354 to 176,348) but decreases slight for quantile regression (AIC from 222,510.3 to 222,510.6). We also notice that significance of affiliation count drops in NB regression (from p$<$0.0001 to p=0.11669) and after we removed affiliation count from the NB model, we noticed no difference in model quality (AIC remains at 176,348) but the significance of country count increases (from p=0.146 to p$<$0.0001).

\textbf{The coefficients of the final regression model including author-related factors} are shown in Tables \ref{tab:c_nb_age_team} and \ref{tab:c_quan_age_team}. With respect to author teams, we find that both models show that author team's past citations and author count increase the number of citations. Both models also suggest that an increase in the number of past papers by the author team has a negative impact on citations. Finally, both models also suggest that a larger number of affiliations (affiliating organizations or countries) in a paper increases citations. In NB regression, this is shown with the positive relationship between number of author countries and citations count. The number of author countries is not significant in quantile regression but we think the same effect is captured by author affiliation count that is a positive predictor. We observed a Pearson correlation of 0.64 between affiliation count and country count  supporting this idea.

\textbf{Author team vs. paper venues:} We think it is interesting to compare whether author team properties offer a better model of citations than venue when adjusting for age. When using the same data with the same articles removed due to missing author team information, AIC scores show that in NB regression, venue is a better predictor than author team by having lower AIC 174,491 vs 176,348. Similarly in the quantile regression model, the venue is a better predictor by having lower a AIC value: 221,688 vs. 222,510.

\begin{table}
  \caption{Coefficients of negative binomial regression for age and author team}
  \label{tab:c_nb_age_team}
\centering
\begin{tabular}{lrrrr}
  \hline
 & Estimate & Std. Error & z value & Pr($>$$|$z$|$) \\ 
  \hline
(Intercept) & -1.1851 & 0.0564 & -21.01 & 0.0000 \\ 
  Age & -0.0658 & 0.0021 & -31.86 & 0.0000 \\ 
  log(Age) & 1.7579 & 0.0233 & 75.46 & 0.0000 \\ 
  log(PastCites) & 0.2515 & 0.0120 & 20.89 & 0.0000 \\ 
  log(PastPapers) & -0.1386 & 0.0191 & -7.26 & 0.0000 \\ 
  log(Authors) & 0.1937 & 0.0221 & 8.76 & 0.0000 \\ 
  Countries & 0.0768 & 0.0170 & 4.51 & 0.0000 \\ 
   \hline
\end{tabular}
\end{table}

\begin{table}
  \caption{Coefficients of quantile regression for age and author team}
  \label{tab:c_quan_age_team}

\begin{tabular}{lrrrr}
  \hline
                   & Value    & Std. Error & t value   & Pr($>$$|$t$|$) \\
  \hline
(Intercept)        & -8.8848 & 0.4918     & -18.0664 & 0.0000                \\
Age                & -0.3881 & 0.0159     & -24.3478 & 0.0000                \\
log(Age)           & 7.0966  & 0.1588     & 44.6833  & 0.0000                \\
log(PastCites) & 1.2917  & 0.1190     & 10.8539  & 0.0000                \\
log(PastPapers)  & -0.6446 & 0.1829     & -3.5242  & 0.0004                \\
log(Authors)  & 2.3605  & 0.2557     & 9.2306   & 0.0000                \\
Affiliations  & 0.3138  & 0.1104     & 2.8429   & 0.0045                   \\
  \hline
\end{tabular}
\end{table}

\subsection{RQ4: Impact of paper-related factors on citations}
\label{sec:rq4}
\textbf{Paper (venue) Type}
Scopus provides multiple paper (venue) types. In our dataset, the different types and number of items of each type were: Article (20,881),  Article in Press (438), Conference Paper (1,328), Conference Review (13), Editorial (1,192),  Erratum (82), Letter (237), Note (208), Review (710), and Short Survey (23). As we had extracted only journal papers, we were not expecting  conference papers  to appear in our dataset. We checked some of those papers and found that those papers were journal extensions of conference papers. For simplified data analysis, we merged groups Article, Article in Press and Conference paper to one a group called Article. Review and Short Surveys were merged into one group called Review. We left Editorials as their own group and we grouped all the rest (Conference Review, Erratum, Letter, and Note) in a group called Others. We found that Editorials and Other papers receive significantly (p$<$0.0001) fewer citations than Articles. This is true for both NB and quantile regression. However, only in NB regression (predicting mean citations). Review papers receive significantly more citations than Articles (p$<$0.0001) in quantile regression the impact of reviews is not statistically significant (p=0.44). We found that after adding paper type to the baseline model, with only age, the model quality of the cases of both NB regression (AIC from 186,298 to 184,642) and quantile regression (AIC from 235,909 to 235,195)  improved.

\textbf{Paper length} should in principle increase citations as longer papers should have more contributions and therefore be more frequently cited, an observation that has been reported in the literature \cite{ahmed2016impact, fox2016citations}. Furthermore, Editorials are typically quite short so their lower citation rate might be explained by paper length. At this point, we removed 910 articles with missing page counts. Again we found that models get more improvement if we take a log of page count instead of the raw value. In NB regression page count is highly significant (p$<$0.0001) positive predictor of citation count and model quality improves (AIC from 180448 to 178,945). In quantile regression, page count is also highly significant positive predictor and model quality improves (AIC from 227783 to 227427). Furthermore, when page count is included publication type Review becomes significant positive predictor also in Quantile regression (from p=0.44 to p=0.014).

\textbf{The number of references in a paper} have also been shown to be a positive predictor of citations \cite{fox2016citations}. This is most likely due to the fact that better articles that attract more citations are more likely to also have done a more thorough review of literature. We measured the number of references as well as their recency. Recency is measured with the Price index \cite{price1970citation}, i.e. the share of references that have been published in the past five years. 
Both numbers of references and Price index are highly significant (p$<$0.0001) in NB regression and model quality improves (AIC from 178,945 to 177,210). In quantile regression, the same holds as both are highly significant (p$<$0.0001) and model quality also improves (AIC from 227,427 to 226,832).  

\textbf{Title Length} of papers has also shown to be a significant negative predictor of citations in past works \cite{letchford2015advantage, Jamali2011, guo2018succinct, BRAMOULLE2018311} 
Since the information was available, we investigated whether this is true in SE articles as well. We decided to test both measuring titles in words and in characters. We find that shorter title measured both in character and in words are significant predictors. However, they are highly correlated and when both of them are added to the same NB regression model only the number of words is a significant predictor of citations (p=0.00186). The highest improvement in model quality is achieved when only the number of words is retained then the number of words becomes highly significant (p$<$0.0001) and model quality improves (AIC from 177,210 to 177,128). In quantile regression, neither words (p=0.071) nor characters (p=0.257) improve the model quality. 

Coefficients of the \textbf{final models including all the paper-related factors} are shown in Tables \ref{tab:c_nb_age_paper} and \ref{tab:c_quan_age_paper}. We can see that both models are highly similar. In both models, publications of types Editorial or Other are associated with lower citation count when compared to regular research articles. On the other hand, publications of type 'Reviews' are associated with more citations in comparison to regular research articles. We can see that an increase in page count increases citation count, however, log transformation of the page count is performed suggesting diminishing returns when articles get longer and longer. With respect to references, we can see that both increasing in the absolute number of reference and increasing the ratio of recent references measured with Price index is associated with higher citation count. The only difference between the models is that in NB regression, shorter title lengths, measured in words, is associated with higher citation count. For quantile regression, the relationship between title length and citation count does not improve model quality (AIC) and is not statistically significant.    

\textbf{Paper vs Venue}. When comparing model quality, we found that, in both NB (AIC Paper: 177,128 vs. AIC Venue: 178,716) and quantile regression models (AIC paper: 226,832 AIC Venue: 226881), paper-related factors have better AIC than venue factors when adjusted for age. Thus, from the three candidates paper-related factors, venue, and author-related factors, paper factors comes on top. Of course, this should not be a surprise. 

\begin{table}
 \caption{Coefficients of negative binomial regression for age and paper properties}
  \label{tab:c_nb_age_paper}
\centering
\begin{tabular}{lrrrr}
  \hline
 & Estimate & Std. Error & z value & Pr($>$$|$z$|$) \\ 
  \hline
(Intercept) & -1.6593 & 0.0604 & -27.48 & 0.0000 \\ 
  Age & -0.0852 & 0.0019 & -43.75 & 0.0000 \\ 
  log(Age) & 1.7785 & 0.0245 & 72.70 & 0.0000 \\ 
  Type:Editorial & -1.0408 & 0.0564 & -18.46 & 0.0000 \\ 
  Type:Other & -1.2279 & 0.0746 & -16.47 & 0.0000 \\ 
  Type:Review & 0.2878 & 0.0551 & 5.22 & 0.0000 \\ 
  log(Pages) & 0.4139 & 0.0157 & 26.41 & 0.0000 \\ 
  Price & 1.1086 & 0.0319 & 34.70 & 0.0000 \\ 
  References & 0.0170 & 0.0007 & 25.99 & 0.0000 \\ 
  TitleLength & -0.0261 & 0.0028 & -9.43 & 0.0000 \\ 
   \hline
\end{tabular}
\end{table}

\begin{table}
  \caption{Coefficients of quantile regression for age and paper properties}
  \label{tab:c_quan_age_paper}
\begin{tabular}{lrrrr}
  \hline
                 & Value   & Std. Error & t value  & Pr($>$$|$t$|$) \\
                   \hline
(Intercept)       & -9.6191 & 0.4226     & -22.7596 & 0.0000                \\
Age               & -0.4750 & 0.0149     & -31.9454 & 0.0000                \\
log(Age)          & 7.2635  & 0.1840     & 39.4718  & 0.0000                \\
Type:Editorial & -2.0658 & 0.2753     & -7.5042  & 0.0000                \\
Type:Other     & -2.6035 & 0.3448     & -7.5513  & 0.0000                \\
Type:Review    & 2.0416  & 0.6345     & 3.2179   & 0.0013                \\
log(Pages)    & 1.3838  & 0.1355     & 10.2162  & 0.0000                \\
Price             & 5.0603  & 0.3215     & 15.7417  & 0.0000                \\
References       & 0.1431  & 0.0102     & 14.0021  & 0.0000     \\
  \hline
\end{tabular}
\end{table}

\subsection{RQ1-4: Complete regression model including all factors}
\label{sec:RQ5_completeModel}

Here we combine individual models presented with RQ1-RQ4. Complete models for NB and quantile regression are in Tables \ref{tab:c_nb_all} and \ref{tab:c_quan_all}. The variables that we include in the complete model have been studied in RQ1-RQ4 and presented in Tables \ref{tab:c_bnr_age_veneu}-\ref{tab:c_quan_age_paper}. Venue was modelled in both with each venue as dummy variable or as using impact factor scores of each venue in Section \ref{sec:ResRq2}. 
However, using them simultaneously is not possible when computing regression models. 
Both venues as dummy variables, see Tables \ref{tab:c_nb_all} and \ref{tab:c_quan_all}  and as impact factors scores are shown, see Tables \ref{tab:c_nb_all_vif} and \ref{tab:c_quan_all_vif}. 
The text follows venues as dummy variables models as it offered better model quality. 

\begin{table}
 \caption{Coefficients of negative binomial regression. Venues as dummies}
  \label{tab:c_nb_all}
\centering
\begin{tabular}{lrrrr}
  \hline
 & Estimate & Std. Error & z value & Pr($>$$|$z$|$) \\ 
  \hline
(Intercept) & -3.8814 & 0.1042 & -37.24 & 0.0000 \\ 
  Age & -0.0875 & 0.0021 & -41.71 & 0.0000 \\ 
  log(Age) & 1.7994 & 0.0235 & 76.43 & 0.0000 \\ 
  \hdashline
  J:ESE & 0.6451 & 0.0793 & 8.13 & 0.0000 \\ 
  J:IJSEKE & -0.6919 & 0.0771 & -8.97 & 0.0000 \\ 
  J:ISSE & 0.1658 & 0.0999 & 1.66 & 0.0970 \\ 
  J:IST & 0.8091 & 0.0698 & 11.60 & 0.0000 \\ 
  J:JSEP & 0.1839 & 0.0830 & 2.21 & 0.0268 \\ 
  J:JSS & 0.8512 & 0.0683 & 12.46 & 0.0000 \\ 
  J:JSTT & 0.5620 & 0.0809 & 6.95 & 0.0000 \\ 
  J:REJ & 0.8794 & 0.0918 & 9.58 & 0.0000 \\ 
  J:SOSYM & 0.6614 & 0.0825 & 8.02 & 0.0000 \\ 
  J:SQJ & 0.2639 & 0.0832 & 3.17 & 0.0015 \\ 
  J:STVR & 0.5369 & 0.0927 & 5.80 & 0.0000 \\ 
  J:SW & 1.6925 & 0.0733 & 23.10 & 0.0000 \\ 
  J:SWPE & 0.2585 & 0.0689 & 3.75 & 0.0002 \\ 
  J:TOSEM & 1.1699 & 0.1093 & 10.71 & 0.0000 \\ 
  J:TSE & 1.8442 & 0.0691 & 26.68 & 0.0000 \\ 
   \hdashline
  log(PastCites) & 0.1417 & 0.0110 & 12.82 & 0.0000 \\ 
  log(PastPapers) & -0.0841 & 0.0174 & -4.83 & 0.0000 \\ 
  log(Authors) & 0.0171 & 0.0208 & 0.82 & 0.4099 \\ 
  Countries & 0.0890 & 0.0156 & 5.70 & 0.0000 \\ 
   \hdashline
  Type:Editorial & -0.6941 & 0.0660 & -10.52 & 0.0000 \\ 
  Type:Other & -1.0414 & 0.0866 & -12.03 & 0.0000 \\ 
  Type:Review & 0.3098 & 0.0529 & 5.85 & 0.0000 \\ 
  log(Pages) & 0.7176 & 0.0188 & 38.14 & 0.0000 \\ 
  Price & 0.9226 & 0.0306 & 30.19 & 0.0000 \\ 
  References & 0.0120 & 0.0006 & 19.22 & 0.0000 \\ 
  TitleLength & -0.0151 & 0.0027 & -5.58 & 0.0000 \\ 
   \hline
\end{tabular}
\end{table}

\begin{table}
 \caption{Coefficients of complete negative binomial regression. Venues as impact factor}
  \label{tab:c_nb_all_vif}
\centering
\begin{tabular}{lrrrr}
  \hline
 & Estimate & Std. Error & z value & Pr($>$$|$z$|$) \\ 
  \hline
(Intercept) & -3.0655 & 0.0698 & -43.93 & 0.0000 \\ 
  Age & -0.1008 & 0.0020 & -50.18 & 0.0000 \\ 
  log(Age) & 1.8663 & 0.0234 & 79.71 & 0.0000 \\ 
    \hdashline
  CiteScore\_a & 0.0062 & 0.0131 & 0.47 & 0.6369 \\ 
  SNIP\_a & 0.5238 & 0.0164 & 31.98 & 0.0000 \\ 
    \hdashline
  log(PastCites) & 0.1552 & 0.0111 & 13.94 & 0.0000 \\ 
  log(PastPapers) & -0.1213 & 0.0176 & -6.91 & 0.0000 \\ 
  log(Authors) & 0.0485 & 0.0212 & 2.29 & 0.0219 \\ 
  Countries & 0.0894 & 0.0159 & 5.63 & 0.0000 \\ 
    \hdashline
  Type:Editorial & -0.9569 & 0.0615 & -15.55 & 0.0000 \\ 
  Type:Other & -1.3600 & 0.0865 & -15.73 & 0.0000 \\ 
  Type:Review & 0.3909 & 0.0538 & 7.27 & 0.0000 \\ 
  log(Pages) & 0.3881 & 0.0158 & 24.63 & 0.0000 \\ 
  Price & 0.9805 & 0.0310 & 31.61 & 0.0000 \\ 
  References & 0.0117 & 0.0006 & 18.50 & 0.0000 \\ 
  TitleLength & -0.0237 & 0.0027 & -8.72 & 0.0000 \\ 
   \hline
\end{tabular}
\end{table}

\begin{table}
 \caption{Coefficients of complete quantile regression.  Venues as dummies}
  \label{tab:c_quan_all}
\begin{tabular}{lrrrr}
  \hline
&Value                & Std. Error & t value & Pr($>$$|$t$|$)         \\
  \hline
(Intercept)          & -17.8471 & 1.0988 & -16.2421 & 0.0000 \\
Age                  & -0.4471  & 0.0218 & -20.5196 & 0.0000 \\
log(Age)             & 7.6221   & 0.2532 & 30.1070  & 0.0000 \\
 \hdashline
J:ESE    & 2.7842   & 1.0614 & 2.6233   & 0.0087 \\
J:IJSEKE & -3.2079  & 0.8212 & -3.9066  & 0.0001 \\
J:ISSE   & -0.4250  & 0.9051 & -0.4695  & 0.6387 \\
J:IST    & 2.4506   & 0.7692 & 3.1860   & 0.0014 \\
J:JSEP   & 0.0361   & 0.9794 & 0.0369   & 0.9706 \\
J:JSS    & 2.1154   & 0.7398 & 2.8594   & 0.0043 \\
J:JSTT   & 1.1157   & 0.8578 & 1.3006   & 0.1934 \\
J:REJ    & 4.5804   & 1.6590 & 2.7610   & 0.0058 \\
J:SOSYM  & 2.5169   & 0.8656 & 2.9077   & 0.0036 \\
J:SQJ    & 1.5076   & 0.8540 & 1.7652   & 0.0775 \\
J:STVR   & 1.0547   & 1.0857 & 0.9715   & 0.3313 \\
J:SW     & 6.2616   & 0.8352 & 7.4971   & 0.0000 \\
J:SWPE   & 0.0066   & 0.7383 & 0.0089   & 0.9929 \\
J:TOSEM  & 33.9346  & 8.0838 & 4.1979   & 0.0000 \\
J:TSE    & 14.4304  & 1.1610 & 12.4293  & 0.0000 \\
 \hdashline
log(PastCites)   & 0.8554   & 0.1333 & 6.4170   & 0.0000 \\

log(PastPapers)    & -0.6066  & 0.1927 & -3.1488  & 0.0016 \\
log(Authors)    & 0.9041   & 0.2443 & 3.7005   & 0.0002 \\
Affiliations    & 0.3824   & 0.1191 & 3.2122   & 0.0013 \\
 \hdashline
Type:Editorial    & -0.4070  & 0.5637 & -0.7221  & 0.4703 \\
Type:Other        & -3.1333  & 0.6691 & -4.6833  & 0.0000 \\
Type:Review       & 1.8818   & 1.0723 & 1.7549   & 0.0793 \\
log(Pages)       & 2.2930   & 0.2052 & 11.1752  & 0.0000 \\
Price                & 4.4699   & 0.3437 & 13.0056  & 0.0000 \\
References          & 0.1005   & 0.0113 & 8.8805   & 0.0000 \\
  \hline
\end{tabular}
\end{table}

\begin{table}
 \caption{Coefficients of complete quantile regression.  Venues as impact factor.}
  \label{tab:c_quan_all_vif}
\begin{tabular}{lrrrr}
  \hline
                  & Value    & Std. Error & t value  & Pr($>$$|$t$|$) \\
                    \hline
(Intercept)        & -23.5381 & 0.7397     & -31.8217 & 0.0000                \\
Age                & -0.4609  & 0.0190     & -24.2988 & 0.0000                \\
log(Age)           & 7.9125   & 0.2187     & 36.1877  & 0.0000                \\
 \hdashline
CiteScore\_a       & 0.5898   & 0.1551     & 3.8037   & 0.0001                \\
SNIP\_a            & 9.1705   & 0.7813     & 11.7377  & 0.0000                \\
SJR\_a             & -15.4454 & 2.5651     & -6.0214  & 0.0000                \\
 \hdashline
log(PastCites) & 0.9399   & 0.1176     & 7.9897   & 0.0000                \\
log(PastPapers)  & -0.7418  & 0.1798     & -4.1248  & 0.0000                \\
log(Authors)  & 0.9391   & 0.2347     & 4.0011   & 0.0001                \\
Affiliations  & 0.4716   & 0.1137     & 4.1462   & 0.0000                \\
 \hdashline
Type:Editorial  & -1.3882  & 0.4644     & -2.9896  & 0.0028                \\
Type:Other      & -3.2742  & 0.6899     & -4.7458  & 0.0000                \\
Type:Review     & 2.2631   & 0.9919     & 2.2817   & 0.0225                \\
log(Pages)     & 1.9671   & 0.1677     & 11.7316  & 0.0000                \\
Price              & 4.6065   & 0.3428     & 13.4390  & 0.0000                \\
References        & 0.1078   & 0.0101     & 10.6755  & 0.0000               \\
  \hline
\end{tabular}
\end{table}

The complete models for most parts give similar results as individual models (RQ1-RQ4). Next, we outline the differences. With respect to \textbf{paper age}, we can see that the coefficients of individual and complete negative binomial regression are similar, as shown in Tables \ref{tab:nbr_coef_age} and  \ref{tab:c_nb_all}. This means that age is a coefficient that is not largely influenced by other coefficients. The same is true for quantile regression strengthening the results, see Tables \ref{tab:nbr_coef_age} and  \ref{tab:c_nb_all}. 

\textbf{Venue.} We modelled venues in two ways as dummy variables, Tables \ref{tab:c_nb_all} and \ref{tab:c_quan_all} and as impact factor scores, as shown in Tables \ref{tab:c_nb_all_vif} and \ref{tab:c_quan_all_vif}.
Comparisons of the tables show that other predictors are not largely affected by this modeling choice. We can see that there is a somewhat large change in page count.
However, we think this is due to the IEEE SW effect that we discuss below. Overall our results suggest that for future studies either modeling choice is appropriate. The text that follows is based on the venues as dummy variables. 
When we look at the publication venue factor, we can see differences between the individual and complete models. Individual model for negative binomial regression, Table \ref{tab:c_bnr_age_veneu}, shows highest positive coefficients for TSE (1.37), TOSEM (1.07), and ESE (0.6). However, when we look at the complete model, \ref{tab:c_nb_all}, we can see that the three journals with the highest positive coefficients are TSE (1.84), IEEESW (1.69), and TOSEM (1.17). In the individual model, the IEEE SW magazine only had a coefficient of 0.27. We think this radical change in the coefficient between individual and complete model is caused by journal guidelines of IEEE SW that restrict the number of pages and references. In the complete model, an increase in both page and reference count are associated with an increase in expected citations. Once the number of pages and references are controlled for in the regression model journal, IEEE SW becomes associated with a high expected number of citations. In the complete model of quantile regression, the journals with the three largest positive coefficients are TOSEM (33.9), TSE (14.4), and IEEE SW (6.2). This supports the finding that IEEE SW is highly cited venue after we control for page count and references. We can also see that in the complete quantile regression mode, TOSEM has clearly the highest coefficient, which contradicts the individual model, as seen in  Table \ref{tab:c_quan_age_veneu}. However, in the complete model, the standard error value of TOSEM is clearly the largest among quantile regression coefficients. 
This suggests the possibility of 'noisy' data for TOSEM in Scopus.  We find that complete model predictions of TOSEM are based on only 199 papers while individual model prediction in RQ1 was based on 468 articles. We found that, in the complete model, 263 of 268 removed TOSEM articles have been excluded due to missing page count in Scopus data. Our further investigation found that this affects the newest articles published 2010 onward. Thus, with respect to TOSEM, we conclude that the set of articles in the complete model is small and biased towards older articles making the coefficient of TOSEM volatile. 

\textbf{Author-related factors.} When we look at the author related factors (RQ3) in the individual and complete model, we can see that the effects of past citation, past papers, and size of author team get smaller in negative binomial regression see Tables\ref{tab:c_nb_age_team} and \ref{tab:c_nb_all}. In fact, the size of the author team is not significant in the complete model (p=0.4) However, country count coefficient increases in the complete model. Similar phenomena happens with quantile regression as effects of past citation, past papers, and size of author team get smaller, see \ref{tab:c_quan_age_team} and \ref{tab:c_quan_all}. Likewise in quantile regression, affiliation count that measure networking effect has a higher coefficient in the complete model. We think that the positive effects of the past citation and author team size that can be seen as a proxy of author team quality are partially captured by other variables in the complete model. For example, a high-quality author team is more likely to aim for venues with higher prestige and perform better-related work resulting in higher Price Index. However, it seems the other variables in the complete cannot capture the networking effects of having authors from multiple affiliations or countries involved. 

\textbf{Paper-related Factors.} With respect to paper factors (RQ4), we found that effect of paper type (Article, Editorial, Review, or Other) is similar in both individual and complete negative binomial regression, see Tables\ref{tab:c_nb_age_paper} and \ref{tab:c_nb_all}. Reviews are associated with higher citation count than regular Articles while Editorials and Others are associated with lower citation count. In the complete model with all factors, the negative association of Editorial (from -1.04 to -0.69) and Other (from -1.23 to -1.04) is smaller than in the individual model. For Reviews, there is a mild increase in the coefficient (from 0.29 to 0.31). In Quantile regression, the difference between Article and Editorial decreases as Editorials coefficients gets closer to zero (from -2.06  to -0.41) and is no longer significant in the complete model (p=0.47). The coefficient of Reviews gets smaller in the complete quantile regression model (from 2.04 to 1.88). On the other hand, the negative coefficient of Other gets larger (from -2.6 to -3.1). Overall, it appears that the negative association of Editorials to citation decreases when page length is taken into consideration in the complete model. 

For quantitative paper-related factors in negative binomial regression, we see that coefficient of page count increases (from 0.41 to 0.71) when moving from individual to complete model. This is in line with the fact that negative effect of shorter paper types Editorial and Other decreases in the complete model. The page count coefficient also increases in quantile regression from (1.38 to 2.29). The coefficient of Price index decreases slightly both in negative binomial (from 1.11 to 0.92) and quantile regression (from 5.06 to 4.47) when moving from individual to complete model. Same is true for the number of references in the paper in negative binomial (from 0.017 to 0.012) and quantile (from 0.14 to 0.10) regression models. The title length was only part of negative binomial regression and we find that its negative impact decreases (from -0.026 to -0.015) but is still highly significant.

\textbf{Example.}Finally, as an illustrative example, we present the expected number of citations with the complete models of negative binomial and quantile regression. We set the dummy variables as follows, the type of the publication is set to Article and journal is set to JSS as those two groups are the largest in our dataset. We fix other factors to their median values that along with 5, 25, 75, and 95 percentiles are shown in Table \ref{tab:quantiles}, e.g. median for age is 14 in the complete model data. Table \ref{tab:c_nb_pred} show the expected mean citation of a paper when factors are between 5 and 95 percentile.


For example, when the paper-age factor is in 5 percentile we expect 2.21 citations when other factors are fixed to their medians and dichotomous factor Paper type is Article and Journal is JSS. Table \ref{tab:c_quan_pred} shows the expected median citations with Quantile regression. From \ref{tab:quantiles}, we can see that 5 percentile of age means two or that 95 percentile of past citations means 2,477.4.


We see that author, affiliation and country have quite limited effect due to their limited range between 5 to 95 percentile 1-5, 1-3, and 1-2 respectively. On the other hand, we can see how the past citation count increases excepted citations from 12.96 to 39.21 when moving from 5 percentile to 95 percentile in negative binomial regression. Other factors whose 5 to 95 percentile change impact to the expected mean citation is more than 10 citations in negative binomial regression are the past articles, page count, Price index, and the number of references. 
Finally, we should mention that Tables \ref{tab:c_nb_pred} and \ref{tab:c_quan_pred} contain situations that are unreal. For example, 5 percentile of the past papers means that the team has previously published 0 papers. Yet, as the other factors are fixed to their median, this means that this same author team would have 124 past citations. Obviously having past citations without past articles is not possible. So, these tables should be looked with caution and they are presented only as examples of the practical impact of our factors to the expected number of citations.

\begin{table}[hbt]
\caption{Expected citation of negative binomial regression when factors are in 5, 25, 50, 75, 95 percentile and other factors are fixed to the median (50 percentile), and publication type is Article and journal is JSS}
  \label{tab:c_nb_pred}
\centering
\begin{tabular}{lrrrrr}
  \hline
factor & 5\% & 25\% & 50\% & 75\% & 95\% \\ 
  \hline
Age & 2.21 & 11.26 & 25.68 & 28.24 & 19.74 \\ 
  PastCites & 12.96 & 19.36 & 25.68 & 31.76 & 39.21 \\ 
  PastPapers & 35.24 & 28.40 & 25.68 & 23.81 & 21.92 \\ 
  Authors & 25.38 & 25.68 & 25.68 & 25.86 & 26.09 \\ 
  Countries & 25.68 & 25.68 & 25.68 & 25.68 & 28.07 \\ 
  Pages & 6.36 & 18.70 & 25.68 & 33.17 & 48.55 \\ 
  Price & 17.53 & 21.09 & 25.68 & 31.21 & 44.11 \\ 
  References & 22.25 & 23.34 & 25.68 & 29.65 & 39.52 \\ 
  TitleLength & 27.69 & 26.47 & 25.68 & 24.92 & 23.46 \\ 
   \hline
\end{tabular}
\end{table}

\begin{table}[hbt]
\caption{Expected citation of quantile regression when factors are in 5, 25, 50, 75, 95 percentile and other factors are fixed to the median (50 percentile), and publication type is Article and journal is JSS}
  \label{tab:c_quan_pred}
\centering
\begin{tabular}{lrrrrr}
  \hline
factor & 5\% & 25\% & 50\% & 75\% & 95\% \\ 
  \hline
Age & 0.62 & 7.21 & 10.09 & 9.73 & 7.21 \\ 
  PastCites & 5.96 & 8.38 & 10.09 & 11.37 & 12.64 \\ 
  PastPapers & 12.37 & 10.81 & 10.09 & 9.54 & 8.95 \\ 
  Authors & 9.46 & 10.09 & 10.09 & 10.46 & 10.92 \\ 
  Affiliations & 10.09 & 10.09 & 10.09 & 10.47 & 11.24 \\ 
  Pages & 5.63 & 9.08 & 10.09 & 10.91 & 12.12 \\ 
  Price & 8.24 & 9.13 & 10.09 & 11.03 & 12.71 \\ 
  References & 8.88 & 9.29 & 10.09 & 11.30 & 13.71 \\ 
     \hline
\end{tabular}
\end{table}

\begin{table}[hbt]
\caption{Percentiles 5, 25, 50, 75 and 95 of our data used for the complete model (n=22,773)}
  \label{tab:quantiles}
\centering
\begin{tabular}{lrrrrr}
  \hline
 & 5\% & 25\% & 50\% & 75\% & 95\% \\ 
  \hline
Age & 2.00 & 6.00 & 14.00 & 24.00 & 37.00 \\ 
  PastCites & 0.00 & 16.00 & 124.00 & 559.00 & 2477.40 \\ 
  PastPapers & 0.00 & 12.00 & 42.00 & 105.00 & 281.40 \\ 
  Authors & 1.00 & 2.00 & 2.00 & 3.00 & 5.00 \\ 
  Countries & 1.00 & 1.00 & 1.00 & 1.00 & 2.00 \\ 
  Affiliations & 1.00 & 1.00 & 1.00 & 2.00 & 4.00 \\ 
  Pages & 2.00 & 9.00 & 14.00 & 20.00 & 34.00 \\ 
  Price & 0.00 & 0.20 & 0.41 & 0.62 & 1.00 \\ 
  References & 0.00 & 4.00 & 12.00 & 24.00 & 48.00 \\ 
  TitleLength & 4.00 & 7.00 & 9.00 & 11.00 & 15.00 \\ 
   \hline
\end{tabular}
\end{table}






\section{Discussion}
\label{sec:discussion}
We discuss our results with respect to prior works, as discussed in Section \ref{sec:Related_Work}. We give special attention to two past papers \cite{Tahamtan2016, Onodera2015} that have provided comprehensive reviews of the research area. Tahamtan et al. \cite{Tahamtan2016} provided a larger review but the drawback is that review also included many papers that use univariate analysis yielding for weaker evidence. Onodera  \cite{Onodera2015} reviewed a smaller set of studies (14 paper in total) and also provided results from their own analysis. We found the findings of Onodera et al. to be of higher quality, since all its reviewed studies used robust multivariate analysis such as multiple regression. Table \ref{tab:summary} summarizes our results and compares it to prior works. In second (\cite{Tahamtan2016}), third (\cite{Onodera2015}), and fourth (TP) column of Table \ref{tab:summary}, the first number shows the amount of support between the row factor and citation count. 
The second number is amount of no support for row factor and citation count. In the columns for papers \cite{Tahamtan2016} and \cite{Onodera2015}, the numbers refer to number of studies supporting or not supporting. For this paper (column heading TP), it is the number of models offering support. This paper has two individual models and two complete models so four in total are used for comparison. From Tahamtan et al. \cite{Tahamtan2016}, we counted the supporting and non-supporting papers from in-text citations. Let us now acknowledge that there may be inaccuracies due to ambiguities in the written text of \cite{Tahamtan2016}. From Onodera et al. \cite{Onodera2015}, we counted supporting / non-supporting evidence from their Table 15, which made our task easy. 

\begin{table*}[hbt]
\caption{Summarizing and comparing our findings to two previous studies \cite{Tahamtan2016, Onodera2015}. TP = This paper. First number is the amount positive support between the row factor and citation count. The second number is amount of no support or the negative relationship between row factor and citation count. In past papers \cite{Tahamtan2016, Onodera2015}, the numbers in the cells refer to number of studies. For this paper, it is the number of models offering support. 
}
  \label{tab:summary}
\begin{tabular}{llllm{10.5cm}}
\hline
                    & \cite{Tahamtan2016} & \cite{Onodera2015} & TP & TP Notes                                                  \\
                    \hline
Age                 &22/0   & NA  & 4/0  & Relationship is not linear.   \\
Venue Impact            &32/3   &10/0  &  4/0   &                                                        \\
Author \#        &50/8   & 10/4   &  3/1   &                                                        \\
Past Citations    &10/3   & 6/0  & 4/0    &                                                        \\
Past Papers    &12/1   & 3/1  & 0/4   & Past Papers is always negative in presence of Past Citations.                                                         \\
Aff or Country \#    &18/1   & 2/3   & 4/0   & Either affilication or country count was significant but not both                                                       \\
Review Paper       & 7/0  & NA  & 3/1    &                                                        \\
Paper Length        & 22/3   & 5/7   &4/0    &                              \\
Reference \#         & 17/0  & 9/0  & 4/0      &                                                  \\
Reference Recency   & 1/0  & 4/0   &4/0    &                                                        \\
Title Length        & NA  & NA   & 2/2   & Shorter titles are prefered (measured as number of words)                                                      \\
\hline
\end{tabular}
\end{table*}

\subsection{RQ1: Age}
Multiple papers have studied the age effect, see Tahamtan et al. \cite{Tahamtan2016}. They pointed out that, in general, citation rate is faster in the first years and citation rates go down as time passes. However, there are considerable differences between fields. Lynn \cite{lynn2014diffusing} studied six different disciplines and showed that, in Cell biology, highest (absolute not age adjusted) mean citation counts by large margin are in articles with age 3-5 years. While, in Sociology, the highest mean citation counts are in articles with age 18-20 years. From Figure \ref{fig:CitesPerAge}, we can see that peak mean citation count in SE occurs in 17 years. In this respect, SE citation practices are closer to sociology were citation occur during longer time period than to cell biology that is characterized by rapid burst of citations right after the publication of the paper.

Age can be controlled with multiple ways: (1) by studying citation of articles to a particular year only, e.g. \cite{Onodera2015}; (2) Dividing citations by the paper age \cite{garousi2017quantity}; (3) By using paper age as a a singe predictor in the regression model \cite{sin2011international}; and (4) by modelling age as two predictors, e.g., age and $age^2$, to allow non-linear relationship between age and citation count \cite{ayres2000determinants}. We chose the option (4) above and modelled age with two predictors, $age$ and $log(age)$. 
We also tried age and $age^2$ but found better model quality with the former. We chose option (4) as it allows the most accurate modelling of age without removing any data.  

If we compare the individual age model, Tables \ref{tab:nbr_coef_age} and \ref{tab:quant_coef_age} and the complete model, Tables \ref{tab:c_nb_all} and \ref{tab:c_quan_all}, we can see that age coefficients in our models are not heavily influenced by the presence of other variable. This means that age is a robust predictor of citations when modelled with two parameters as we have done. 

\subsection{RQ2: Venue}
Past literature, see Table \ref{tab:summary}, shows strong support for the idea that impact factors have a positive relationship with expected citations. Our findings support this. Furthermore, we compared tree metrics from Scopus (SNIP, SJR, CiteScore) and the choice of whether to use yearly or average values of impact factors. Our results partially favour simpler metric Citation Score, which is Scopus' implementation of the Web-of-Science Impact Factor, instead of more complex SNIP\cite{moed2010measuring} and SJR\cite{gonzalez2010new}. Surprisingly, we found better results when using average impact factor instead of using yearly impacts. This suggest that following impact factor over multiple years is better than using the latest one when planning on where to publish. 

We found that better model quality (AIC) was achieved when modelling each venue individually than using dummy variables for each venue (dummy variables) than when using any of the impact factor measures. Naturally, modelling each venue individually has the drawback that it does not scale very well. 

In a related work, Garousi and Fernandez \cite{garousi2016citations} showed with Scopus data that TOSEM papers had an average more citation than TSE or ESE papers. They reported average citations for TOSEM, TSE, and ESE: 47.4, 36.8, and 18.4. When we compute the averages, we find for TOSEM, TSE and ESE the values of 51.4, 54.7 and 21.8, respectively. Thus, it seems that citation counts to these three journals have been increasing overall and that TSE has become more cited than TOSEM. In their paper \cite{garousi2016citations}, the authors also adjusted for age by dividing the citations by paper age and reported average age 'normalized' citations for TOSEM, TSE, ESE as follows: 3.9, 2.5, and 2.2. When we computed the age normalized averages, we find for TOSEM, TSE and ESE the values of 3.9, 3.1, and 2.9, respectively. This shows that TSE and ESE citation counts have been increasing while TOSEM has remained the same. Garousi and Fernandez \cite{garousi2016citations} had data set ending in 2014 while our data set ended in 2018. Thus, we can say that this change has occurred in the past 4 years (2014-2018). 
Finally, we need to point out that our data shows TSE as being the top-cited paper when age adjustment is done by allowing none-linear relationship between citation count and paper age, see Figure \ref{fig:CitesPerJournalPerYearNB}. We think modelling for non-linear relationship is essential as early years result in more citation than later years.  

When we compare the individual venue model, Section \ref{sec:ResRq2} and the complete model, Section \ref{sec:RQ5_completeModel}, we can see that IEEE Software has a higher positive association in the complete model. IEEE Software has for years enforced strict page and reference count limits. This results in paper of having less pages and less references than papers in other outlets. Higher page and reference count are associated with higher citation count in our complete models. Once those factors are included in the complete model IEEE Software rises among the top three outlets with respect to expected citation count. 

One maybe wondering whether one should look coefficients of individual venue models or the complete models when trying figure out which are the top cited venues in software engineering. There is no clear cut answer for this. One should first answer whether one cares about the venue citations only or is one interested in the venue citations after other factors such as page count are controlled for.  

\subsection{RQ3: Author-related factors}
Multiple author related factors have been studied. Past work shown in Table \ref{tab:summary} shows support to the idea that the more authors in the paper is associated with higher citation. However, there is also considerable counter evidence.  In our paper, higher author count was associated with higher citation count in both individual models. However, in the complete model predicting for mean citation author count was no longer statistically significant (p=0.41). In Tables \ref{tab:c_quan_pred} and \ref{tab:c_nb_pred}, we can see small impact of author count on expected mean and median citations when moving between 5-95 percentile in our data, 25.38-26.09(mean) 9.46-10.95 (median) citations.

Table \ref{tab:summary} supports the idea that authors' past citation count are associated with future citation. We found a clear and strong relationship between author team's past citation and the paper citations. Many of the past works have considered authors individually \cite{hurley2013deconstructing} or author pairs \cite{tang2014scholars}, while we studied the entire author team. 

Authors past productivity (measured with number of papers) is linked to higher citation counts, see Table \ref{tab:summary}. If we consider productivity without citation count we find this to be true in our data as well. However, past productivity measured in paper count and citations are highly correlated (Pearson r=0.75). When we consider them jointly,  we find the opposite. In other words, 
authors team's productivity has negative relationship with citation count when author team's past citations are also part of the model. This goes against on what has been proposed in prior studies. 

With respect to network effects, Tahamtan et al.  \cite{Tahamtan2016} founds strong support for it while cites only one study for counter evidence \cite{didegah2013factors}. However, in the work by Onodera et al. \cite{Onodera2015} that only reviewed higher quality studies there is more weight on the counter evidence, see Table \ref{tab:summary}. Our results only partly support the network effects. In negative binomial regression model number of countries was a significant positive predictor while the number of affiliating organization was not. In quantile regression that models median expected citations it was the opposite as number of organization was a significant while number of countries was not. These two variables are correlated in our data (Pearson r=0.67).  Tables \ref{tab:c_quan_pred} and \ref{tab:c_nb_pred} show small impact of affiliation or country count on expected mean and median citations when moving between 5-95 percentile in our data, 25.68-28.07(mean) 10.09-11.24 (median) citations. 

\subsection{RQ4: Paper-related factors}
Tahamtan et al. \cite{Tahamtan2016} cited seven papers that support the idea of review papers receiving more citations while citing no papers that did not find the relationship. Onodera et al. \cite{Onodera2015} did not investigate this topic. However, our results support this idea. In all four models (two paper-related models and two complete models), we found that review paper type has positive association with citation count and in 3 out of 4 models this is also statistically significant (p$<$0.05). Garousi and Fernandez \cite{garousi2017quantity} showed previously that reviews had more average and median citations. Also in our other previous work \cite{garousi2016systematic}, in which we had conducted a systematic literature review of literature reviews in software testing, we had found that, both in terms of both citation metrics (total number and average annual number of citations), citations to the secondary studies (review papers) are higher than primary studies papers in two areas of software testing (i.e., web testing and GUI testing) 


Tahamtan et al.  \cite{Tahamtan2016} showed strong support for the positive relationship between citation count and number of pages in a paper with 22 papers for and 3 against, see Table \ref{tab:summary}. Onodera et al. \cite{Onodera2015} provided a more conflicting picture with 5 papers supporting while 7 paper rejecting this hypothesis.
All of our four models support this with high statistical significance (p$<$0.0001). In our models, page count receives the highest change in expected citation with NB regression when moving from 5 percentile to 95 percentile in our data, 6.36-48.55 see Table \ref{tab:c_nb_pred}. We speculate that past works might have failed to find evidence of page count if they have not factored journal specific limits to their model. In practice, this happens by entering each journal as predictor rather than using a an impact factor as proxy. For example in our complete models we can see decrease in page count coefficient when replacing individual journals with their impact factors in NB regression (from 0.72 to 0.39) and Quantile regression (from 2.29 to 1.96), see Tables \ref{tab:c_nb_all}-\ref{tab:c_quan_all_vif}.

Tahamtan et al.  \cite{Tahamtan2016} cited 17 papers that support the positive relationship between citation count and number of references in a paper. Onodera et al. \cite{Onodera2015} also offered strong support for this with 9 studies in favour and non against. We found this to be true in all four models  with high statistical significance (p$<$0.0001). 

With respect to recency of reference, e.g. the Price index, being positively associated with citations Tahamtan et al. are not very specific and only indicate one study supporting this. Onodera et al. \cite{Onodera2015} offered four studies on the topic and all supported the hypothesis. All our models offered strong support for Price index, i.e. the share of references that are from past five years, having positive association with citation count. When price index varies 5-95 percentiles in our data, it means between zero and one, see Table \ref{tab:quantiles}. This results in big changes in expected mean and median citation, 17.53-44.11 and 8.24-12.71 respectively. 

Finally, we looked into whether title length affects citation counts. The idea that short tiles receive more citations was popularized few years ago by Nature News article \cite{deng2015papers}. A few possible reasons for this phenomena have been reported, e.g., "tightly constructed titles are in authors’ best interests", says Karl Ziemelis, Nature’s chief physical sciences editor, as "they can improve the chances of a paper being discovered by readers and encourage them not to pass it over" \cite{deng2015papers}. The story was based on \cite{letchford2015advantage} that analyzed 140,000 papers from Scopus. Counter evidence from small scale study of 150 articles shows  that in medicine that longer titles receive more citations \cite{jacques2010impact}. Yearly variation is addressed in \cite{guo2018succinct} that uses 300,000 economic papers and shows that title length is negatively correlated in year 1956-2000 but becomes positive after year 2000. The authors speculate that is due to online searching. Yet, using the same economic paper data but with more controls concludes that shorter titles are better even controlling for yearly variation \cite{BRAMOULLE2018311}. 

In our dataset, we found that shorter titles are associated with more citations both in quantile and NB regression. However, they are statistically significant (level p=0.05) only in NB regression that predicts mean citations. However, the impact of title in NB regression is stronger than author or country count. In other words, moving from 4 word title (5 percentile) to 15 word title (95 percentile) reduces expected mean citation from 27.69 to 23.46, see Table \ref{tab:c_nb_pred}. 

\subsection{RQ5: Expected mean and median}
Recently there has been an increase in the number of studies that have used quantile regression to study citations
\cite{danell2011can,wang2018relationship, ahlgren2018exploring, sienkiewicz2016impact, stegehuis2015predicting}. However, only one compares quantile regression with other regression approaches \cite{wang2018relationship}. Wang \cite{wang2018relationship} computes several quantiles and finds that quantile regression finds higher ``positive relationship between the number of editorial board members and the citations per paper for the 30'' than ordinary least-squares regression (OLS). 

We computed the median quantile only. We did this as we think it might be more important to understand the factors of median expected citations than mean expected citations. However, when we look at the predictors of the complete models for both NB regression and quantile regression, Tables \ref{tab:c_nb_all} and \ref{tab:c_quan_all}, we find that there are not very big differences and they do not affect the most impactful predictors, such as the author team's past performance, age, page count, reference count and recency of references. 

We think the differences between mean and median citations stem from the very high citation count of top-cited papers that greatly affect mean expected citations. We can see that with respect estimates on journals coefficients there are minor differences but as some journals are rather small this also reflected in the higher error terms.  We can see that the author count is not statistically significant when predicting mean citations. We find that title length was a negative predictor of only mean citations.  We also found that when predicting mean country count is significant but when predicting median this is replaced with author affiliation count. Impact of review paper type is less significant in median citations. 

\section{Limitations}

Any issues in Scopus data are also problems in our dataset.  We tried our best to clear out most of the  blunt errors, e.g. in Section \ref{sec:ResRq2}, we explained how we removed authors whose profile included works from multiple authors. Another example is that TOSEM journal papers where we had to remove many papers in the complete models due to missing page counts. This resulted in high error terms in complete model. In general, we recommend investigating the error terms in addition to statistical significance. Overall, many past works have used Scopus data and we are not aware of any other major deviations than authors whose profile included works from multiple authors which affects Chinese authors in particular \cite{ioannidis2018thousands}. However, it is still possible to that that using another bibliometric database such as Web of Science could bring different results. 


\section{Conclusions}
\label{sec:Conclusions}
In this paper, we have investigated the factors affecting citations to papers published in software engineering (SE) journals. We collected data from the Scopus database. We utilized two regression models. First, negative binomial regression models were used to investigate mean citations. Second, due to the skewed nature of citation data, we investigated median citations with quantile regression.  

Our main findings are the followings. First, factors affecting citations in SE journals are to a large degree similar to prior works from other fields, see Table \ref{tab:summary}. Most important factors explaining citations are age, venue, author team past performance, paper length and reference count and recency (Price Index \cite{price1970citation}). However, we found a notable difference in comparison to prior works as we showed that author team productivity, measured with the number of past papers, is negatively associated with citations when author team past citations are part of the model. Thus, the author team with short but highly cited publication list is associated with higher citation count. Second, with respect to past SE literature, we state that this is the first paper that uses robust multivariate models to investigate how multiple factors impact papers citations. 

Given that we now have the state of the art knowledge on factors influencing citations, we offer some advice to researchers: 
(1) Aim for venue that has high coefficient in the complete models, see Tables \ref{tab:c_nb_all} and \ref{tab:c_quan_all}. But let us note that our study sheds no light whether the citations that papers in particular venue receive are due to paper quality or halo effect of a particular venue \cite{johnson1997getting, hudson2007known}. 
(2) Build a high-quality author team with highly-cited but rather short publication list. However, again we cannot say whether this is due to author quality or halo effect. 

(3) Aim for high-quality work that has comprehensive content (thus longer paper length). Papers with more reference and more pages are associated with higher citation count. Recency of references is also important predictor and we think this is a proxy on how well the author knows the most recent prior work. Finally, finding a proper and concise' (short) title  may help your citation count.

Many interesting factors affecting citation were not covered in this paper due to Scopus not having such data or for the required manual analysis that was not possible with our large 25,000+ paper dataset. Our comparisons to  Tahamtan et al.  \cite{Tahamtan2016} and Onodera et al. \cite{Onodera2015} in Table \ref{tab:summary} show that many of our findings are a good match with previous studies in other areas of science. Thus, while we would wait for future bibliometric studies in SE, like the current one, to address the missing factors, one can speculate that the findings of Tahamtan et al.'s review \cite{Tahamtan2016} and our study can act as a substitute in the meanwhile.

\bibliographystyle{ACM-Reference-Format}
\bibliography{999_Bib} 
\end{document}